\documentstyle[aps,epsf]{revtex}
\begin{document}
\draft

\title{
 Theory of Drop Formation
    }
\author{
Jens Eggers
 \\Universit\"at Gesamthochschule Essen, Fachbereich Physik, \\ 
  45117 Essen, Germany}
\maketitle
\begin{abstract}
We consider the motion of an axisymmetric column of Navier-Stokes
fluid with a free surface. Due to surface tension, the thickness of
the fluid neck goes to zero in finite time. After the 
singularity, the fluid consists of two halves,
which constitute a unique continuation of the 
Navier-Stokes equation through the singular point.
We calculate the asymptotic
solutions of the Navier-Stokes equation, both before and after
the singularity. The solutions have scaling form, characterized 
by universal exponents as well as universal scaling functions,
which we compute without adjustable parameters. 
\end{abstract}
\newpage
\section{Introduction}
The breakup of free-surface flows has been an object of intense
research from the advent of hydrodynamic theory, and in particular
the discovery of surface tension \cite{/L/,/S/,/H/}.
Namely, surface tension is the driving force behind this 
phenomenon, as it tends to reduce the surface area by decreasing 
the radius of a column of fluid. This indeed leads to the 
formation of drops, as is seen most clearly from Rayleigh's 
\cite{/R/} stability analysis of an infinite cylinder of fluid
with radius $r_0$. 

He considered perturbations of different wavelengths and calculated 
their growth rates. While long wavelength perturbations result in
the smallest surface area, they require large mass transport
between maxima and minima. Both effects strike a balance at 
the wavelength $\lambda \approx 9r_0$, corresponding to the fastest 
growing mode. This type of analysis subsequently has been greatly refined, 
for example including viscosity \cite{/W/}, surface charges \cite{/GM/},
or higher order nonlinear effects \cite{/CR/}.

However, even higher order perturbation theory rapidly becomes 
inadequate as the thickness of the fluid neck goes to zero at a 
point, and fluid is expelled from this region with increasingly 
high speed. Near the singularity, characterized by a blow-up of
local curvature, and of the velocity at the pinch-point, nonlinear
effects will soon dominate the dynamics. An asymptotic 
scaling theory of this singularity, where surface tension, viscous, 
and inertial forces are balanced, has been presented very recently
\cite{/E/}. 

But eventually the size of the neck or the times scale on which it 
is moving will reach microscopic scales, and a hydrodynamic 
description breaks down altogether. For example, the neck will
evaporate somewhere close to the pinch point, where it has minimum 
thickness. Shortly after that, new surfaces will have formed on 
either side, and this time the dynamics is described by two 
separate Navier-Stokes problems. The physical question we address 
here is whether the new initial conditions depend on the microscopic
mechanisms behind the breakup. In other words, taking two different
kinds of fluids with the same surface tension, density, and viscosity, 
will the breakup look the same on scales larger than the microscopic
ones?

We will indeed show that drop formation is a hydrodynamic phenomenon
in the above sense. Namely, we construct asymptotic solutions to the 
Navier-Stokes equation after breakup, which describe two separate 
surfaces and which are unique continuations of the solutions before
breakup. The physical origin of this uniqueness lies in the
properties of the solution before breakup \cite{/E/}. 
The diameter of the fluid neck does not go to zero uniformly,
but only inside a ``hot'' region around the pinch point. 
Outside, the solution is static on the time scale of the central region.
As one approaches the singularity, the size of the hot region
goes to zero. Hence by the time microscopic mechanisms
become important, their action is confined to an extremely small 
region in space. The continuation is achieved by matching the
outer parts of the solution before breakup onto 
the corresponding regions after breakup. Since the outer 
parts are virtually unaffected by the microscopic dynamics,
this procedure yields universal continuations.  
  
This seems to be the first example of a partial differential 
equation uniquely describing a ``topological transition'' \cite{/GPS/}. 
The result is also important for numerical simulations, 
which usually rely on some ad-hoc prescription for the 
formation of a new surface\cite{/TSO/,/LNSZZ/}, or for breakup in
related physical situations \cite{/CDGKSZ/}. 

Our paper is organized as follows: In Section 2 we derive a 
one-dimensional approximation of the Navier-Stokes equation 
\cite{/BFL/,/ED/}, valid as the ratio $\epsilon$ of the radial to 
the longitudinal scale of the flow is small. They have self-
similar pinching solutions, which are described by a pair of 
scaling functions $\phi(\xi)$ and $\psi(\xi)$ for the radius of the 
fluid neck
and the velocity, respectively. As the time distance from the
singularity goes to zero, the slenderness parameter $\epsilon$
for this solution vanishes, making it an exact solution 
asymptotically.

The scaling functions $\phi$ and $\psi$ obey two pairs of 
ordinary differential equations, one for the time before breakup,
the other for the time after breakup. For most of the rest of this 
paper, we will be constructing unique solutions to those equations.
In the third section we consider the similarity equations 
before breakup. Shortly before the singularity, the fluid far 
outside the pinch region is no longer able to follow the motion
near the pinch point. This leads to boundary conditions for the 
similarity functions at infinity, and together with a regularity 
condition in the interior,
a unique solution of the equations is selected. We compute
this solution numerically.

The same procedure is adopted in the fourth section for the similarity
equations after breakup. Here the solution is matched onto the 
profiles before breakup. This solution
has two halves, each of which is fixed uniquely by the matching.

The concluding discussion gives an example for the breakup of 
a real fluid, which could be measured experimentally. We also
supply numerical evidence for the uniqueness and stability of
our theoretical predictions, and discuss related work.

\section{Similarity equations}

Let us begin by formulating the Navier-Stokes problem for
an axisymmetric column of fluid, where we assume 
the azimuthal velocity to be zero. A sketch of the geometry of
the problem can be found in Figure 1. For a fluid with kinematic
viscosity $\nu$, surface tension $\gamma$, and density $\rho$
the Navier-Stokes equation reads in cylindrical coordinates \cite{/LL/}:

\begin{equation}\label{(2.1)}
\partial_t v_r + v_r\partial_rv_r + v_z\partial_zv_r = -\partial_r p/\rho
\\
  + \nu(\partial^{2}_{r}v_r + \partial^{2}_{z}v_r + \partial_rv_r/r - v_r/r^2),
\end{equation}

\begin{equation}\label{(2.2)}
\partial_t v_z + v_r\partial_rv_z + v_z\partial_zv_z =
-\partial_z p/\rho \\
 + \nu(\partial^{2}_{r}v_z + \partial^{2}_zv_z + \partial_rv_z/r) - g,
\end{equation}
with the continuity equation 

\begin{equation}\label{(2.3)}
\partial_rv_r + \partial_z v_z + v_r/r = 0 .
\end{equation}
The acceleration of gravity points in negative z-direction. 
Here $v_z$ is the velocity along the axis, $v_r$ the velocity
in the radial direction, and $p$ the pressure. There are two
boundary conditions, coming from the balance of normal forces,

\begin{equation}\label{(2.4)}
{\bf n\; \sigma\; n} = -\gamma(1/R_1 + 1/R_2) ,
\end{equation}
and tangential forces

\begin{equation}\label{(2.5)}
{\bf n} \; {\bf \sigma} \; {\bf t} = 0 .
\end{equation}

In (\ref{(2.4)}),(\ref{(2.5)}) we denoted the outward normal
and tangent vector to the surface by ${\bf n}$ and ${\bf t}$,
${\bf \sigma}$ is the stress tensor, and $(1/R_1+1/R_2)/2$
the mean curvature. A standard formula for bodies of revolution
gives

\begin{equation}\label{(2.6)}
\frac{1}{R_1} + \frac{1}{R_2} = \frac{1}{H (1 + (\partial_z H)^2)^{1/2}} - \\
\frac{\partial_z^2H}{(1 + (\partial_z H)^2)^{3/2}} , 
\end{equation} 
where $H(z,t)$ is the radius of the fluid neck, as seen in Figure 1.
The equation of motion for $H(z,t)$ is  

\begin{equation}\label{(2.7)}
\partial_t H + v_z\partial_z H = v_r|_{r=H},
\end{equation}
which says that the surface moves with the fluid at the boundary.

Equations (\ref{(2.1)})-(\ref{(2.7)}) constitute a complex 
moving boundary value problem, which we want to investigate 
near a singularity, where nonlinear effects are bound to
become dominant. The reason exact solutions, valid arbitrarily
close to the singularity, can nevertheless be found, is that 
only very few terms in the equations contribute to the leading
order force balance. Thus to proceed, we first have to identify
those leading order terms. We will then construct explicit 
solutions to the leading order equations and demonstrate their 
consistency with both the internal structure of the Navier-
Stokes equation and with boundary conditions. 

The relevant terms are identified using two properties of the 
singularity to be validated later:

\begin{itemize}

\item[(i)]
The singularity is line-like, i. e. its axial extension 
is much greater than its radial extension.

\item[(ii)]
Surface tension, viscous, and inertial forces are equally 
important near the singularity.

\end{itemize}

Conditions (i) and (ii) are now incorporated into a perturbation
theory. According to (i) we will 
assume that the motion of the fluid at a given time is 
characterized by an axial length scale $\ell_z$ and a radial 
length scale $\ell_r$, for which

\begin{equation}\label{(2.8)}
\ell_r = \epsilon \ell_z,
\end{equation}
where $\epsilon $ is some small parameter. The physical meaning
of $\epsilon$ will come out later from the description of the
singularity. Also introducing a time scale $t_z$ of the singularity, we can 
nondimensionalize all quantities according to

\begin{eqnarray}\label{(2.9)}
&& r=\ell_r \tilde{r} \quad,\quad z=\ell_z \tilde{z} 
\quad,\quad t=t_z \tilde{t} \quad, \\
\nonumber
&& H=\ell_r \tilde{H} \quad,\quad {\bf v}=\frac{\ell_z}{t_z} \tilde{{\bf v}}
\quad,\quad \frac{p}{\rho}=\frac{\ell_z^2}{t_z^2} 
\frac{\tilde{p}}{\tilde{\rho}} \quad, \\ 
\nonumber
&& \nu=\frac{\ell_z^2}{t_z}\epsilon^n \tilde{\nu} \quad,\quad 
\frac{\gamma}{\rho}=\frac{\ell_z^3}{t_z^2} \epsilon^m 
\frac{\tilde{\gamma}}{\tilde{\rho}}
\quad,\quad g=\frac{l_z}{t_z^2} \epsilon^l \tilde{g} \quad. \\
\nonumber
\end{eqnarray}

The scales $\ell_z$, $\ell_r$, and $t_z$ are defined to be constants,
so their derivative with respect to time is zero. However, one must 
bear in mind that the characteristic scales of the singularity change, 
so $\ell_z$, $\ell_r$, and $t_z$ will be different in different 
stages of the singularity formation.
Since there are two length scales $\ell_z$ and $\ell_r$, there is
a certain freedom in the nondimensionalization of the material 
parameters $\nu$, $\gamma/\rho$, and $g$. This freedom is 
completely specified by the exponents $n$, $m$, and $l$ in
(\ref{(2.9)}). We will see below that the exponents are fixed
by the requirement (ii). 

Since the radial extension of the fluid is small, we can
expand all fields in the dimensionless radial variable 
$\tilde{r}$ :

\begin{equation}\label{(2.11)}
\tilde{v}_z(\tilde{z},\tilde{r},\tilde{t}) = \sum_{j=0}^{\infty} 
\tilde{v}_{2j}(\tilde{z},\tilde{t}) (\epsilon \tilde{r})^{2j} \quad,
\end{equation}

\begin{equation}\label{(2.12)}
\tilde{v}_r(\tilde{z},\tilde{r},\tilde{t}) = -\sum_{j=0}^{\infty} 
\frac{\tilde{v}_{2j}'(\tilde{z},\tilde{t})}{2j+2} 
(\epsilon \tilde{r})^{2j+1} \quad,
\end{equation}
and

\begin{equation}\label{(2.13)}
\tilde{p}(\tilde{z},\tilde{r},\tilde{t}) = \sum_{j=0}^{\infty} 
\tilde{p}_{2j}(\tilde{z},\tilde{t}) (\epsilon \tilde{r})^{2j} \quad.
\end{equation}

The definition of $\tilde{v}_r$ automatically ensures incompressibility.
We now insert (\ref{(2.9)})-(\ref{(2.13)}) into the equations of
motion (\ref{(2.1)})-(\ref{(2.7)}), and compare powers in $\epsilon$.
The lowest order expressions result in a closed set of 
equations for $\tilde{v}_0$ and $\tilde{H}$,

\begin{equation}\label{(2.14)}
\partial_{\tilde{t}}\tilde{v}_0 + \tilde{v}_0\partial_{\tilde{z}}\tilde{v}_0 =
  -\frac{\tilde{\gamma}}{\tilde{\rho}} \epsilon^{m-1}
\partial_{\tilde{z}}\left(\frac{1}{\tilde{H}}\right) + 3\tilde{\nu}\epsilon^n
\frac{\partial_{\tilde{z}}\left[(\partial_{\tilde{z}}
\tilde{v}_0)\tilde{H}^2\right]}
{\tilde{H}^2} - \tilde{g}\epsilon^l \quad,
\end{equation}

\begin{equation}\label{(2.15)}
\partial_{\tilde{t}} \tilde{H} + \tilde{v}_0 \partial_{\tilde{z}}\tilde{H} = 
 - (\partial_{\tilde{z}}\tilde{v}_0) \tilde{H}/2 \quad.
\end{equation}

To obtain closure at higher orders in $\epsilon$,
one needs to expand each of the coefficients $\tilde{v}_{2j}$ and
$\tilde{p}_{2j}$, as well as $\tilde{H}$  into a separate power 
series in $\epsilon$. There then exists a consistent representation
of (\ref{(2.1)})-(\ref{(2.7)}) to all orders in $\epsilon$ \cite{/F/}.
We will not be concerned with the explicit form of the higher 
order equations here, so for simplicity we use the notation 
$v_0$ and $H$ (or its
nondimensional counterpart) for the lowest order terms in the 
expansion in $\epsilon$. 

It is evident from (\ref{(2.14)}) that the exponents $m$, $n$, and $l$
determine the balance of forces at leading order.
Since the $1/\tilde{H}$ term, which comes from the 
radius of curvature perpendicular to the axis, is driving the 
instability, it must clearly be present and in fact becomes
infinite at the singularity. At the small scales involved in 
singularity formation, viscosity will also be important.
Finally, velocities are expected to blow up as ever smaller 
amounts of liquid are driven by increasingly large pressure gradients.
Hence we also expect inertial effects to be involved 
asymptotically. Since the acceleration of the fluid diverges 
at the pinch point, the constant acceleration of gravity 
will drop out of the problem. This is precisely the assumption
(ii), incorporated by 
choosing $m=1$, $n=0$, and $l>0$ in (\ref{(2.9)}), which 
leads to an equation where surface tension, viscous, and
inertial forces are balanced, while gravity is irrelevant. 
These assumptions will be tested for consistency later.

We now identify the scales involved in the formation of the 
singularity. It is crucial to notice that all {\it external} 
length  and time scales, which are imposed by boundary and initial 
conditions, do not enter the description of the singularity.
In a jet experiment, for example, external scales would be the 
radius of the nozzle and the period of the driving frequency. 

Near the singularity, the length scales characterizing the
solution become arbitrarily small, while time scales become shorter and 
shorter as one approaches the singularity. Hence the singularity
moves on scales widely separated from the external scales.
It is for this reason that for the mathematical analysis of the 
singularity we do not have to make the boundary or initial 
conditions explicit. Boundary and initial conditions will become
important when we describe numerical simulations of real experiments,
which confirm the consistency of our approach.

The proper units in which to represent the motion near the singularity
can thus involve only {\it internal} parameters of the fluid.
This leaves us with the units of length and time

\begin{equation}\label{(2.16)}
\ell_{\nu}=(\rho\nu^2)/\gamma \quad,\quad t_{\nu}=(\rho^2\nu^3)/\gamma^2 \quad.
\end{equation}
Assuming that the singularity occurs at a point $z_0$, and at a time
$t_0$, the space and time distance from the singularity is 
properly measured as 

\begin{eqnarray}\label{(2.20)}
&& z'=(z-z_0)/\ell_{\nu} \quad, \\
\nonumber
&& t'= (t-t_0)/t_{\nu} \quad. \\
\nonumber
\end{eqnarray}
The units $\ell_{\nu}$ and $t_{\nu}$ are a measure of the width
of the critical region, and are fixed for a given fluid. Singular 
behavior is expected for $|z'| \ll 1$ and $|t'| \ll 1$. Note the
conceptual difference to the characteristic scales $\ell_z$, 
$\ell_r$, and $t_z$ of the singularity, which change in time.

In the variables z' and t', the fluid velocity and the neck radius are:

\begin{eqnarray}\label{(2.17)}
v(z',t') \equiv \frac{t_{\nu}}{\ell_{\nu}} v_0(z,t) \quad, \\
\nonumber
h(z',t') \equiv \ell_{\nu}^{-1} H(z,t) \quad. \\
\nonumber
\end{eqnarray}
Keeping the same terms as in (\ref{(2.14)}),(\ref{(2.15)}) with 
$m=1$, $n=0$, and $l>0$, we find in the limit 
$\epsilon \rightarrow 0$
\begin{equation}\label{(2.18)}
\partial_{t'} v + v\partial_{z'}v = -\partial_{z'}\left(\frac{1}{h}\right) + 
3\frac{\partial_{z'}\left[(\partial_{z'}v)h^2\right]}{h^2} \quad,
\end{equation}

\begin{equation}\label{(2.19)}
\partial_{t'} h + v \partial_{z'}h = - (\partial_{z'}v) h/2 \quad.
\end{equation}
All material parameters have dropped out of the equations, since
everything has been expressed in units of $\ell_{\nu}$ and $t_{\nu}$.

At this point it is worthwhile to pause and notice that we have
already succeeded in reducing the original Navier-Stokes problem 
in two spatial dimensions and in time with a moving boundary 
to just a coupled set of equations in one space dimension and time, 
at least for small $\epsilon$. 
Approximations for thin liquid threads of the type described here
have in fact a long history, see \cite{/BFL/} for a (by no means
complete) list of earlier references. However it seems that 
(\ref{(2.18)}),(\ref{(2.19)}), which contain the correct surface
tension, inertial, and viscous terms, were first derived in \cite{/BFL/}.
Another related approach
goes by the name of Cosserat equations, see for example \cite{/B/}.
In all previous work except \cite{/E/} though, the resulting one-
dimensional equations are treated as model equations, whose 
quality of approximation depends on the particular physical 
situation for which they are used. In the present paper, we will 
show that (\ref{(2.18)}), (\ref{(2.19)}) become {\it exact} close
to pinch-off.

To this end we have to identify the parameter $\epsilon$.
From the definitions (\ref{(2.16)}) we find

\begin{equation}\label{(2.21)}
t_z/t_{\nu} = \epsilon^2 \frac{\tilde{\gamma}^2}
{\tilde{\rho}^2\tilde{\nu}^3} \quad,
\quad \ell_z/\ell_{\nu} = \epsilon \frac{\tilde{\gamma}}
{\tilde{\rho}\tilde{\nu}^2} \quad.
\end{equation}
Thus up to constants $\epsilon^2$ is the characteristic time
scale of the singularity, written in units of $t_{\nu}$. But
the only such time scale is the nondimensional time distance 
from the singularity $|t'|$ itself. Hence $|t'|$ serves as
the desired smallness parameter. We introduced the modulus 
of $t'$ here, since we need a measure of the time distance
before and {\it after} the singularity. 
As $|t'| \rightarrow 0$, 
all higher order terms vanish and only the leading order 
equations (\ref{(2.18)}),(\ref{(2.19)}) remain. By the same
token, the axial and radial length scales behave like
$\ell_z \sim \ell_{\nu} |t'|^{1/2}$ and
$\ell_r \sim \ell_{\nu} |t'|$. Thus close to the singularity,
all length scales become arbitrarily small compared to any
external length scale, just as we asserted above.

As a corollary to this absence of any fixed length scale in the 
problem, we expect singular solutions to have the similarity 
form

\begin{eqnarray}\label{(2.22)}
h = |t'|^{\alpha_1} \phi(\xi) \quad, \\
\nonumber
v = |t'|^{\alpha_2} \psi(\xi) \quad, \\
\nonumber
\end{eqnarray}
where the similarity variable $\xi$ is defined as $\xi=z'/|t'|^{\beta}$.
A similar ansatz has been used in \cite{/KM/} for a study of
inviscid flow, but in a different geometry. 

The values of the exponents $\alpha_1$, $\alpha_2$, and $\beta$ 
are inferred immediately from dimensional analysis. Namely
$\ell_r \sim \ell_{\nu}|t'|$ implies $\alpha_1 = 1$, $\ell_z/t_z \sim 
(\ell_{\nu}/t_{\nu})|t'|^{-1/2}$ is a typical velocity scale, 
giving $\alpha_2 = -1/2$,
and $\beta = 1/2$ follows from $\ell_z \sim \ell_{\nu} |t'|^{1/2}$. 
The appearance of fractional powers forces us to use the modulus 
of $t'$ in the scaling laws (\ref{(2.22)}). 
The type of similarity solutions we are going to investigate is thus

\begin{eqnarray}\label{(2.23)}
&& h = |t'| \phi(\xi) \quad, \\
\nonumber
&& v = \pm |t'|^{-1/2} \psi(\xi) \quad, \\
\nonumber
&& \xi = \pm z'/|t'|^{1/2} \quad. \\
\nonumber
\end{eqnarray}
The two different signs take care of identical solutions with
different parity.
The acceleration of the fluid diverges like $|t'|^{-3/2}$,
and surface tension, viscous, and inertial forces are balanced.
Since $|t'|\sim\epsilon^2$ we conclude that the exponent $l$ in 
(\ref{(2.9)}) is $l=3$, which is consistent
with our previous assumptions. 

Inserting (\ref{(2.23)}) into (\ref{(2.18)})
and (\ref{(2.19)}) we find that the asymptotic equations of 
motion indeed have scaling solutions, where the scaling functions
$\phi$ and $\psi$ obey the equations

\begin{equation}\label{(2.24)}
s(\psi/2 + \xi \psi'/2) + \psi \psi' = \phi'/ \phi^2 + 3 \psi'' + 
 6 \psi' \phi'/ \phi \quad, 
\end{equation}

\begin{equation}\label{(2.25)}
s(-\phi + \xi \phi'/2) + \psi\phi' = -\psi'\phi/2 \quad.
\end{equation}
The prime refers to differentiation with respect to $\xi$.
The terms in brackets come from the time derivative, $s=1$
refers to the time before the singularity ($t<t_0$), $s=-1$
to the time after the singularity ($t>t_0$).

Hence close to the singularity, $|t'|\ll 1$ and $|z'|\ll 1$, we 
have further reduced the problem to a set of two {\it ordinary
differential equations}. To find unique solutions of
(\ref{(2.24)}) and (\ref{(2.25)}) we still need to formulate 
appropriate boundary conditions. This and the numerical integration of
(\ref{(2.24)}),(\ref{(2.25)}) will be the subject of the next 
two sections, first for $t<t_0$, and then for $t>t_0$.

\section{Before breakup}

In this section we consider the similarity equations (\ref{(2.24)}),
(\ref{(2.25)}) for $s=1$, i.e. before breakup. Some of the 
calculations relevant for the next section will be done for
general $s$. We show that the similarity equations  have 
precisely one physically 
allowed solution, and compute it. Therefore singular solutions 
are completely universal: once the origins of the space and time 
axes are fixed by specifying $z_0$ and $t_0$, there are no more 
free parameters. The relevant units of length and time are set by
the fluid parameters.

As the similarity equations are of first order in $\phi$ and
of second order in $\psi$, solutions are specified by three 
initial conditions $\phi(\xi_i)$, $\psi(\xi_i)$, and $\psi'(\xi_i)$
at a reference point $\xi_i$. Universality implies that we 
need to find three conditions which uniquely fix the physically 
allowed solution. 

For the first condition, suppose we choose a small region of 
width $\ell_{\nu}\delta$ around the singularity, such that 
$\delta \ll \min(1,L/\ell_{\nu})$, where $L$ characterizes some
outer length scale. For $|z'| \leq \delta$ and $|t'| \ll 1$ we
are well within the critical region of the singularity, and effects 
of the boundaries are negligible. Thus the similarity equations 
(\ref{(2.24)}),(\ref{(2.25)}) apply for $|z'| \approx \delta$ and we 
have $|t'|\phi(\pm\delta |t'|^{-1/2}) \approx h(\pm\delta,t')$.
First we observe that the point $|z'| = \delta$ goes to infinity in 
similarity variables as $|t'| \rightarrow 0$. Second, in this limit
$h$ at $|z'| = \delta$ will not be able to follow the motion of the 
singularity, whose width decreases like $|t'|^{1/2}$, and whose
time scale goes to zero with $|t'|$. Hence $h(\pm \delta, t')$ 
must approach a finite value as $|t'| \rightarrow 0$.
To be consistent with this physical requirement, $\phi(\xi)$ must grow
quadratically as $|\xi|$ goes to infinity. 

Hence we have two conditions on the solutions of (\ref{(2.24)}),(\ref{(2.25)}):

\begin{itemize}
\item[a)] $\phi(\xi)$, $\psi(\xi)$ need to be regular on the real axis 
$\xi\in ]-\infty,+\infty[$.

\item[b)] For $\xi\rightarrow\pm\infty$, $\phi(\xi)/\xi^2$ should approach 
a finite limiting value.
\end{itemize}
Conditions similar to b) have also been employed in \cite{/BBDK/}.
Note that the physical concept behind our argument is inertia,
which prohibits the large amount of fluid far away from the 
singularity to move with the fluid in the skinny pinch region. 

We will now show that the requirements a) and b) completely
determine the solution of the similarity equations. In particular,
we do not have to specify the limiting values of $\phi(\xi)/\xi^2$,
they rather come out of the solution of the problem. This is 
consistent because in our analysis we deal exclusively
with the equations of motion valid close to the singularity.
No input from regions where the expansion is not valid is needed.
Thus boundary or initial conditions can enter the problem only implicitly,
as they determine the position of the singularity $z_0$,$t_0$.

Let us begin by examining the behavior of solutions for 
$\xi\rightarrow\pm\infty$. It is advantageous to first eliminate 
$\phi$ from the problem, leaving us with a third-order equation 
for $\psi$ \cite{/J/}. To this end (\ref{(2.25)}) is written as

\begin{equation}\label{(3.1)}
\phi' = \phi \frac{s - \psi'/2}{ \psi + s\xi/2} .
\end{equation}
On one hand this equation can be used to express $\phi$ in
terms of $\psi$,

\begin{equation}\label{(3.2)}
\phi = \left[ (I'-3\psi'')/K - 6\psi' \right]^{-1} \quad,
\end{equation}
where we have introduced the notation 

\begin{eqnarray}\label{(3.3)}
&& K=(s-\psi'/2)/(\psi+s\xi/2) \quad, \\
\nonumber
&& I=(s\xi\psi+\psi^2)/2 \quad.
\nonumber
\end{eqnarray}
On the other hand, writing $\psi$ as an integral over the kernel $K$
we have

\[
\phi = \phi(\xi_0) \exp\left\{\int_{\xi_0}^{\xi} K(\zeta)d\zeta \right\} \quad.
\]
Inserting this into (\ref{(3.2)}), taking the logarithm, and 
differentiating, we find

\begin{equation}\label{(3.4)}
\psi'''=\frac{K}{3}\left\{(I'/K)'+I'+3\psi''(K'/K^2-3)-6\psi'K\right\},
\end{equation}
which is a single equation just in terms of $\psi$.

Plugging the ansatz $\psi = B\xi^{\alpha}$ into  (\ref{(3.4)}),
one finds the leading order behavior on the right hand side to
be

\[
\frac{2B}{3\xi}\left[\frac{(\alpha+1)^2}{4} \xi^{\alpha} +
\frac{\alpha+1}{2} \xi^{\alpha}\right] \quad.
\]
Thus $\psi$ must decay like $1/\xi$ or $1/\xi^3$ at infinity for the 
terms to cancel. 
In particular, growth of $\psi$ at infinity is 
prohibited, since the ``inertial'' term $I'$ is quadratic in $\psi$,
and would grow faster than any other term in (\ref{(3.4)}).

Thus one is lead to an asymptotic expansion of the form

\begin{equation}\label{(3.5)}
\psi=\frac{1}{\xi} \sum_{i=0}^{\infty} b_i\xi^{-2i} \quad.
\end{equation}
Only odd powers appear, since (\ref{(3.4)}) is invariant under 
the transformation $\xi\rightarrow-\xi$ and $\psi\rightarrow-\psi$.
Using (\ref{(3.2)}) we can calculate the leading behavior of $\phi$
corresponding to (\ref{(3.5)}):

\begin{eqnarray}\label{(3.6)}
&& \phi=a_0\xi^2[1+O(\xi^{-2})] \quad, \\
\nonumber
&& a_0=2/[6b_0-sb_1-b_0^2] \quad. \\
\nonumber
\end{eqnarray}

This means (\ref{(3.5)}) represents precisely the physically relevant 
solutions we are interested in. To further investigate the 
expansion (\ref{(3.5)}), we derive recursion relations for the 
coefficients $b_i$ to arbitrarily high order. The lowest order 
expressions are

\begin{eqnarray}\label{(3.7)}
&& b_2=-\frac{1}{2}[3b_0^3+7sb_0b_1] \quad,\\
\nonumber
&& b_3=\frac{s}{8}[-30b_0^3+9b_0^4-148sb_0b_1-9sb_0^2b_1-10b_1^2
-8b_2(10+3b_0)] \quad.\\
\nonumber
\end{eqnarray}
All $b_i$ are thus determined by just two free coefficients,
$b_0$ and $b_1$, or by virtue of (\ref{(3.6)}), $a_0$ and $b_0$.
However, the expansion (\ref{(3.5)}) is only asymptotic, as for
large $i$ the $b_i$ grow like

\[
b_i\sim(-12)^i i! \quad.
\]

This means for large $\xi$ all solutions of (\ref{(3.4)}) are 
{\it up to exponentially small corrections} given by a two-
parameter family of functions $\psi_{a_0b_0}(\xi)$ \cite{/BO/},
which behave like $1/\xi$ asymptotically. The expansion 
(\ref{(3.5)}) is asymptotic to $\psi_{a_0b_0}(\xi)$ and for 
sufficiently large $\xi$ can be used to compute
$\psi_{a_0b_0}(\xi)$ to any desired accuracy. To understand the
significance of this observation, we have to investigate the 
stability of the functions $\psi_{a_0b_0}(\xi)$.

Doing so turns out to be slightly more convenient in the
original space of initial conditions 
$(\phi(\bar{\xi}),\psi(\bar{\xi}),\psi'(\bar{\xi}))$, where 
$\bar{\xi}\gg 1$ is kept fixed. Denoting by $\phi_{ab}(\xi)$
the function $\phi$ corresponding to $\psi_{ab}(\xi)$, we are
interested in particular in perturbations which carry us out 
of the two-dimensional manifold of initial conditions
$(\phi_{ab}(\bar{\xi}),\psi_{ab}(\bar{\xi}),\psi_{ab}'(\bar{\xi}))$.
Differentiating with respect to $a$ and $b$, we find that to leading 
order in $\bar{\xi}$, $(0,0,1)$ is a vector normal to this manifold.

We now consider small perturbations relative to the solutions 
$\phi_{ab}$,$\psi_{ab}$:

\begin{eqnarray}\label{(3.8)}
&& \phi(\xi)=\phi_{ab}(\xi)(1+\epsilon_1(\xi)) \\
\nonumber
&& \psi(\xi)=\psi_{ab}(\xi)(1+\epsilon_2(\xi)) \\
\nonumber
&& \psi'(\xi)=\psi_{ab}'(\xi)(1+\epsilon_3(\xi)). \\
\nonumber
\end{eqnarray}
The correction $\epsilon_3(\xi)$ describes the behavior of perturbations 
perpendicular to the plane of asymptotic solutions $\phi\sim\xi^2$,
$\psi\sim\xi^{-1}$. Inserting (\ref{(3.8)}) into (\ref{(2.24)}),
(\ref{(2.25)}) and linearizing in the $\epsilon_i$ reveals that 
to leading order $\epsilon_3$ behaves like

\begin{equation}\label{(3.9)}
\epsilon_3(\xi)=\epsilon_3(\bar{\xi})\exp
\left\{\frac{s\bar{\xi}}{6}(\xi-\bar{\xi})\right\} \quad.
\end{equation}

Hence for $s=1$ an arbitrarily small perturbation introduced 
at $\bar{\xi}$ will carry the solution away from the physically 
relevant manifold as $|\xi|\rightarrow\infty$. Only a two-dimensional
manifold of solutions is consistent with $\phi(\xi)/\xi^2\rightarrow
const$ as $\xi$ tends to $+\infty$ or $-\infty$. This means the
requirement b) corresponds to {\it two} constraints on physically 
relevant solutions. Since the equations
are of third order, we need to find one additional constraint
to uniquely fix the allowed solutions. 
It is worth remarking that the unstable 
growth (\ref{(3.9)}) comes from the presence of the viscous 
term $\psi''$ in (\ref{(2.24)}). Hence in a strictly inviscid 
theory no selection would take place. 

To find the third constraint we look at condition a), 
saying that $\phi$, $\psi$ 
be regular. Considering (\ref{(3.1)}) this is a nontrivial condition,
as $\psi$ must be bounded and hence there is a point $\xi_0$ 
with

\begin{equation}\label{(3.10)}
\psi(\xi_0)+\xi_0/2=0 \quad.
\end{equation}
Therefore, since $s=1$, the denominator in (\ref{(3.1)}) will 
vanish at $\xi_0$, leading to a singularity unless the condition  

\begin{equation}\label{(3.11)}
\psi'(\xi_0)=2
\end{equation}
is also met. To explore the corresponding regular solutions,
we expand $\psi$ around $\xi_0$:

\begin{equation}\label{(3.12)}
\psi(\xi)=\sum_{i=0}^{\infty} d_i(\xi-\xi_0)^i \quad.
\end{equation}

The function $\phi$ can again be recovered from (\ref{(3.2)}).
We find

\begin{eqnarray}\label{(3.13)}
&& d_0=-\xi_0/2 \quad,\\
\nonumber
&& d_1=2 \quad,\\
\nonumber
&&d_2=-\frac{5\xi_0}{8(3-1/\phi_0)} \quad,\\
\nonumber
&&d_3=\frac{(104-1656\phi_0)d_2^2/75+6\phi_0}{2-36\phi_0} \quad,\\
\nonumber
\end{eqnarray}
where the first two equations follow from (\ref{(3.10)}) and
(\ref{(3.11)}). Just as in the expansion around $\xi=\pm\infty$,
all coefficients $d_i$ are determined by only two coefficients,
$\xi_0$ and $\phi_0=\phi(\xi_0)$. We verified this statement by
deriving recursion relations for the $d_i$ to arbitrarily high
order. This time the expansion has a finite radius of convergence,
whose value depends on the initial conditions $\xi_0$,$\phi_0$.

It is worthwhile to comment on the physical significance of 
$\xi_0$. The equation of motion for the position $z'_s$ of
a marker on the surface $h$ is

\begin{equation}\label{(3.14)}
\partial_t z'_s(t)=v(z'_s(t),t) \quad.
\end{equation}
Rewriting $z'_s$ in similarity variables, $\xi_s=|t'|^{-1/2}z'_s$,
and measuring time on a logarithmic scale, $s=-\ln |t'|$, we find

\begin{equation}\label{(3.15)}
\partial_s \xi_s(s)=\xi_s/2 + \psi(\xi_s) \quad,
\end{equation}
which is the convection equation in similarity variables. Hence 
at the point $\xi_0$, as defined by (\ref{(3.10)}), a surface
marker on $\phi$ is at rest. Regularity properties on such 
``stagnation'' or ``sonic'' \cite{/NR/} points often play a similar 
role in selection. 

To explicitly compute the unique solution of the similarity equations
before breakup, consistent with a) and b), we proceed as follows:
we choose a pair $(\xi_0,\phi_0)$ and compute the Taylor coefficients
$d_i$ to sufficiently high order. This leaves us with a series 
representation of $\psi$ in a disk around $\xi_0$. From there 
onwards, (\ref{(3.4)}) has to be integrated numerically. Since as
$|\xi|\rightarrow\infty$ solutions must be exponentially close to
a two-parameter family of functions $\psi_{ab}$ which are ``repellent'',
solutions will generically not extend to infinity, but rather end 
up in a singularity at finite $\xi$. Dominant balance \cite{/BO/}
in (\ref{(3.4)}) reveals that those singularities have the leading
behavior $\psi(\xi)\sim(\xi-\bar{\xi})^{-1}$. Only a one-dimensional 
submanifold in $(\xi_0,\phi_0)$ is consistent with the solution 
extending to either $+\infty$ or $-\infty$. The point 
$(\bar{\xi}_0,\bar{\phi}_0)$ where both cross corresponds to the 
unique solution we are interested in. 

Our numerical procedure was to introduce $\xi^+$ and $\xi^-$ as 
the values of $|\xi|$ where $|\psi(\xi)|$ exceeded a certain 
bound as $\xi\rightarrow\infty$ or $\xi\rightarrow-\infty$,
respectively. We then optimized $\xi_0$ and $\phi_0$ to give
maximum values of $\xi^+$ and $\xi^-$. As solutions deviate 
exponentially from $\psi_{ab}$, the ``window'' around 
$(\bar{\xi}_0,\bar{\phi}_0)$, which allows for solutions extending
up to a given $|\xi|$ gets small very rapidly with $|\xi|$.
Thus this method allows for a very accurate determination of
$\bar{\xi}_0$ and $\bar{\phi}_0$. The numerical values we found
are quoted, together with other characteristics of the solution,
in Table 1. These results, with the inclusion of the asymptotic 
expansion (\ref{(3.5)}), now allows us to plot the scaling
functions before breakup, $\phi^+$ and $\psi^+$, in Figure 2. 

As seen in Table 1, the stagnation point $\xi_0$ is extremely
close to the point $\xi_{min}$ where $\phi^+$ is minimum.
This means that in the frame of reference of the surface, 
fluid is expelled on either side of the minimum. From 
$z'_{min}=|t'|^{1/2}\xi_{min}$ one sees that the minimum 
moves with velocity $v_{min}=(\xi_{min}/2)|t'|^{-1/2}$.

To make contact with the qualitative description 
of the singularity given in the Introduction, we
schematically divide the similarity solutions 
into three regions: A central region around the minimum
of size $\xi_{central}$, say, where $\phi$ is almost
constant, and outer regions on either side, where $\phi$ 
is quadratic. In this simplified picture, in physical 
space there is a region of size $\xi_{central} |t'|^{1/2}$
around $z_0$, where the diameter of the neck decreases 
linearly in time, and the velocity diverges like $|t'|^{-1/2}$.
Outside this region, both the thickness of the fluid neck 
and the velocity field are constant. Hence as $|t'|\rightarrow 0$,
at any given point $z \neq z_0$ the solution will become static,
and the singularity only occurs {\it at a point} $z_0$ in space.
In terms of some microscopic length $\ell_{micro}$,
one can estimate (molecular) mechanisms to be important 
in a region of size $\xi_{central}(\ell_{micro}\ell_{\nu})^{1/2}$.  

But perhaps the most striking feature is the extreme asymmetry
of $\phi^+$ and $\psi^+$. Indeed, the values of $a_0^{\pm}$,
describing the amplitude of $\phi^+$ as $\xi\rightarrow\pm\infty$,
differ by almost four orders of magnitude. Intuitively, an 
asymmetric solution is to be expected \cite{/ED/}. Namely,
pressure will be higher in the slender part of the solution, 
pushing fluid over to the right. This will cause the right side 
of the solution to fill up with even more fluid and get steeper.
Eventually, this mechanism is only checked by viscosity. But
this argument does not even give an order-of-magnitude estimate
of $a_0^+/a_0^-$. So clearly there is the need for a fully 
analytical theory of the selection problem, which gives at 
least reasonable estimates for the numbers in Table 1.

Another, perhaps related problem pertains to the uniqueness
of the above solution. In principle, the one-dimensional 
submanifolds corresponding to the correct asymptotic behavior
as $\xi\rightarrow+\infty$ and $\xi\rightarrow-\infty$ could 
have several crossings, giving a discrete family of solutions.
The most reasonable guess for a different form of solution 
would be a symmetric one, which would then be highly unstable, 
since small asymmetries would amplify according to the above 
mechanism. Since $\xi_0=0$ for such a solution, $\phi_0$ would
be the only free parameter, which needs to be consistent with
the behavior at infinity. We carefully looked for solutions of this 
type, but found none. Therefore, to the best of our knowledge, 
there is precisely one possible solution, but for a final word
we must await a rigorous mathematical theory.

\section{After breakup}

We now turn to times $t>t_0$, i.e. after breakup. In terms
of the similarity equations (\ref{(2.24)}),(\ref{(2.25)}) this
means we have to put $s=-1$. But apart from the difference in
the equations, there is a completely new type of problem occurring
now, related to the mathematical description of a receding tip.

To understand this, let us consider the asymptotic equations 
(\ref{(2.18)}),(\ref{(2.19)}), which contain the leading order
terms of the Navier-Stokes equation as the slenderness 
parameter $\epsilon$ goes to zero. But this description breaks down 
as one reaches the tip, which is assumed to be at $z'_{tip}(t)$, 
see Figure 3. Namely, the slenderness assumption means that 
$\partial_{z'} h$ is of order $\epsilon$, while $\partial_{z'} h$ 
actually diverges as $z' \rightarrow z'_{tip}$. Indeed, both
the asymptotic form of the pressure gradient 
$(\partial_{z'} h)/h^2$ and of the viscous term 
$\partial_{z'}[(\partial_{z'}v) h^2]/h^2$ diverge as 
$h \rightarrow 0$ and $\partial_{z'} h \rightarrow \infty$.

On the other hand, the complete Navier-Stokes problem has no
singularities as long as $|t'| > 0$. Surface tension will 
ensure that the gradient of the curvature remains finite.
Hence there is a small region around the 
tip, whose width goes to zero as $\epsilon \rightarrow 0$, 
where higher order terms in the Navier-Stokes equation will 
be important. Its size $\ell_{tip}$ can be estimated by saying 
that the asymptotic equations become valid as $\partial_{z'} h$
becomes of order unity at the edge of this region. Thus,
since $\partial_{z'} h \approx \ell_r/\ell_{tip}$, we have
$\ell_{tip} \approx \ell_{\nu}|t'|$, using the known scaling of the radial 
length scale $\ell_r$ with $|t'|$.

Now we transform to similarity variables $\xi = z'/|t'|^{1/2}$,
where $\xi_{tip} = z'_{tip}/|t'|^{1/2}$ is the position of the
tip. Since the width of the tip region shrinks as $|t'|$, it will
go to zero like $|t'|^{1/2}$ even in $\xi$-variables. In the 
neighborhood of any
$\xi \in ]\xi_{tip},\infty[$ the similarity equations will be 
valid as $|t'| \rightarrow 0$. Thus to capture the leading self-similar
behavior of the Navier-Stokes equation after breakup, one just has to 
find the correct boundary conditions for $\phi$ and $\psi$ at
$\xi_{tip}$. This situation is reminiscent of the boundary 
condition at $\xi = \pm \infty$ {\it before} breakup: For $|t'| \rightarrow
0$ the range of validity of the similarity equations extends to
infinity, so supplying boundary conditions at $\xi = \pm\infty$
suffices to uniquely solve the problem. 

To derive the correct boundary condition, we proceed as follows:
We supplement (\ref{(2.18)}),(\ref{(2.19)})  
with higher order terms in $\epsilon$, which regularize the 
equations at the tip. The corresponding  
similarity equations in $\phi$ and $\psi$ now still contain 
$t'$ as a parameter, but are finite as $\xi\rightarrow\xi_{tip}$.
This means solutions of those equations can be supplemented with 
the natural boundary condition $\phi(\xi_{tip})=0$.

Then we derive a simplified version of the equations valid at the
tip, which we can integrate explicitly, using $\phi(\xi_{tip})=0$ as
a boundary condition. {\it Now} we can take the limit 
$|t'|\rightarrow 0$ or $\epsilon\rightarrow 0$, which leaves us
with the correct boundary conditions for $\phi$ and $\psi$, valid for $t'=0$.
We also show that this result is independent of the particular
regularization we have been using, so the result is unique,
as expected from the above argument.
Once the boundary condition has been found, we can solve the similarity
equations to find a unique solution after breakup. 

No knowledge of the fluid motion in the tip region of size
$\ell_{\nu}|t'|$ is needed to calculate the self-similar part
of the solution It remains an interesting open problem to devise
a method to compute an approximate solution in the tip region.
However, since this region becomes arbitrarily small as 
$|t'| \rightarrow 0$, we will not be concerned with this 
question in the present paper.

To construct a regularized version of (\ref{(2.18)}), we observe 
that it can be generalized in the form \cite{/PC/}:
  
\begin{equation}\label{(4.1)}
\partial_{t'}v + v\partial_{z'}v = -\partial_{z'}p + 
\frac{\partial_{z'}\left[(\partial_{z'}v)D^2\right]}{h^2} \quad,
\end{equation}

\begin{equation}\label{(4.2)}
\partial_{t'} h + v \partial_{z'}h = - (\partial_{z'}v) h/2 \quad,
\end{equation}
with

\[
p=\frac{1}{2h}\left[\frac{\partial E}{\partial h} -
\frac{d}{dz'}\frac{\partial E}{\partial(\partial_{z'}h) }\right] \quad.
\]

Here $E=E(h,\partial_{z'}h)$ is a surface energy and 
$D=D(h,\partial_{z'}h)$ a dissipation
kernel. This nomenclature is motivated by the fact that $\partial_{z'}p$ may
be written as 

\[
\partial_{z'}p=\frac{1}{h^2}\frac{d}{dz'}\left[h^2p +
(\partial_{z'}h)\frac{\partial E}{\partial(\partial_{z'}h)}-E\right],
\]
and hence we have the conservation equation 

\begin{eqnarray}\label{(4.3)}
&& \frac{\partial}{\partial t'}\left[h^2v^2/2+E(h,\partial_{z'}h)\right]=
-((\partial_{z'}v)D)^2- \nonumber \\
&& \frac{\partial}{\partial z'}\left[(v^2/2+p)h^2v -
v(\partial_{z'}v)D^2 + (\partial_{t'}h)
\frac{\partial E}{\partial(\partial_{z'}h)}\right] \quad.
\end{eqnarray}

So apart from a surface term this equation says that the sum of 
kinetic and potential energy decreases with a negative definite
dissipation rate ${\cal D}=-((\partial_{z'}v)D)^2$. In the present context,
(\ref{(4.1)}),(\ref{(4.2)}) are phenomenological equations.
There are certainly other higher order correction terms present 
in the Navier-Stokes equation, which have not been included.
However, the only important point here is that $E$ and $D$ can be chosen 
such as to make the equations finite at the tip.
In \cite{/ED/} we already 
introduced a variant of (\ref{(4.1)}),(\ref{(4.2)}) with

\begin{equation}\label{(4.4)}
E(h,\partial_{z'}h)=2h(1+(\partial_{z'}h)^2)^{1/2} \quad.
\end{equation}
This energy is proportional to the surface area and
arizes naturally when keeping the complete curvature 
term in the boundary condition (\ref{(2.4)}).

If the surface at the tip is non-degenerate and the velocity 
field is regular, we simply have $h(z,t)=h_0(t)(z'-z'_{tip})^{1/2}
+O(z'-z'_{tip})^{3/2}$ and 
$v(z,t)=v_0(t)+v_1(t)(z'-z'_{tip})+O(z'-z'_{tip})^2$.
As is verified by inspection, the particular form (\ref{(4.4)})
of $E$ succeeds in keeping $\partial_{z'}p$ finite as $z'\rightarrow z'_{tip}$.
Introducing

\begin{equation}\label{(4.5)}
D(h,\partial_{z'}h)=h(3/(1+(\partial_{z'}h)^2))^{1/2} 
\end{equation}
for the dissipation kernel, the same is true for 
$\partial_{z'}[(\partial_{z'}v)D^2]/h^2$,
hence {\it all} terms in (\ref{(4.1)}) are now finite at the tip.
At the same time, the asymptotic equations (\ref{(2.18)}),(\ref{(2.19)})
are recovered for $\epsilon\rightarrow 0$, as this corresponds to

\begin{eqnarray}\label{(4.6)}
&& E_{asymp}=2h, \\
\nonumber
&& D_{asymp}=\surd 3h .\\
\nonumber
\end{eqnarray}

By construction, all allowed functions $E$ and $D$ must have 
the same limit (\ref{(4.6)}).
We now insert (\ref{(2.23)}) into the regularized equations 
(\ref{(4.1)}),(\ref{(4.2)}). For $t>t_0$, denoting $|t'|^{1/2}$
by $\ell$, we obtain:

\begin{equation}\label{(4.7)}
-\psi/2 - \xi \psi'/2 + \psi \psi' = -G'/ \phi^2 + (\psi'D^2)'/\phi^2
\end{equation}
and

\begin{equation}\label{(4.8)}
\phi - \xi \phi'/2 + \psi\phi' = -\psi'\phi/2 \quad,
\end{equation}
where 

\begin{eqnarray}\label{(4.9)}
&& G=-\frac{\phi}{(1+\ell^2\phi'^2)^{1/2}}-
\ell^2\frac{\phi^2\phi''}{(1+\ell^2\phi'^2)^{3/2}} \quad, \\
\nonumber
&& D=\phi\left(\frac{3}{1+\ell^2\phi'^2}\right)^{1/2} \quad. \\
\nonumber
\end{eqnarray}

Here for simplicity
we have used the special forms (\ref{(4.4)}) and (\ref{(4.5)})
for $E$ and $D$. For $\ell=0$ we recover the asymptotic equations 
(\ref{(2.24)}),(\ref{(2.25)}), while for finite $\ell$ all terms
are regular at the tip as $\phi$ and $\psi$ behave like

\begin{equation}\label{(4.10)}
\phi \sim (\xi-\xi_{tip})^{1/2} \quad,
\psi \sim (\xi-\xi_{tip}) \quad.
\end{equation}

To focus on the tip region, we introduce the rescaled fields
$\bar{\phi}$ and $\bar{\psi}$:

\begin{eqnarray}\label{(4.11)}
&& \bar{\phi}(\zeta)=\phi(\ell\zeta+\xi_{tip}) \quad,\\
\nonumber
&&\bar{\psi}(\zeta)=\ell^{-1}[\psi(\ell\zeta+\xi_{tip})-\xi_{tip}/2] \quad,\\
\nonumber
&& \zeta=\ell^{-1}(\xi-\xi_{tip}) \quad.
\end{eqnarray}
In rescaled variables, the equations are

\begin{equation}\label{(4.12)}
\ell^2\bar{\phi}^2\left[-\bar{\psi}/2-\zeta\bar{\psi}'/2
+\bar{\psi}\bar{\psi}'\right] - \ell\bar{\phi}^2\xi_{tip}/4 = 
\left\{-\bar{G}+\bar{\psi}'\bar{D}^2\right\}' \quad,
\end{equation}
and

\begin{equation}\label{(4.13)}
\bar{\phi} - \zeta \bar{\phi}'/2 + \bar{\psi}\bar{\phi}' = 
-\bar{\psi}'\bar{\phi}/2 \quad,
\end{equation}
with

\begin{eqnarray}\label{(4.14)}
&& \bar{G}=-\frac{\bar{\phi}}{(1+\bar{\phi}'^2)^{1/2}}-
\frac{\bar{\phi}^2\bar{\phi}''}{(1+\bar{\phi}'^2)^{3/2}} \quad, \\
\nonumber
&& \bar{D}=\bar{\phi}\left(\frac{3}{1+\bar{\phi}'^2}\right)^{1/2} \quad. \\
\nonumber
\end{eqnarray}
In (\ref{(4.12)})-(\ref{(4.14)}) and (\ref{(4.15)}) below, primes
refer to differentiation with respect to the rescaled variable $\zeta$.

The only place where $\ell$ still appears is in front of the 
``inertial'' terms on the left hand side of (\ref{(4.12)}).
This is because any fixed region $\zeta \in [0,\zeta_1]$ near the tip 
shrinks to zero in $\xi$-variables as $\ell\rightarrow 0$.
But the fluid at the tip should move with the boundary,
so it is at rest in the frame of reference of the tip. Indeed,
since the left hand side of (\ref{(4.12)}) only contains lower 
order derivatives, the limit $\ell\rightarrow 0$ is regular 
at fixed initial conditions for $\bar{\phi},\bar{\psi}$ at $0$.

Hence by putting $\ell=0$ in (\ref{(4.12)}) we obtain a simplified 
description of the tip region, which is uniformly valid in any fixed
interval $[0,\zeta_1]$. Note that implicitly $\ell$ is still present
by virtue of (\ref{(4.11)}).
Solutions of the resulting equations correspond 
to a very much blown-up version of the tip. Since the solutions are 
regular at $\zeta=0$, we can employ the natural boundary condition
$\bar{\phi}(0)=0$, and from (\ref{(4.14)}) we have 
$\bar{G}(0)=\bar{D}(0)=0$. This means (\ref{(4.12)}) can be
integrated to give

\begin{equation}\label{(4.15)}
\bar{G}=\bar{\psi}'\bar{D}^2 \quad.
\end{equation}

We now supply appropriate matching conditions, which express the
consistency of (\ref{(4.13)}) and (\ref{(4.15)}), valid at the
tip, with the solutions outside the tip. At fixed $\xi$, 
$\phi(\xi)$ and $\psi(\xi)$ are finite in the limit $\ell\rightarrow 0$.
Since the tip region gets arbitrarily small in $\xi$-variables
this is physically reasonable, but has also been checked 
numerically by integrating (\ref{(4.7)}),(\ref{(4.8)}).
Hence we have to require $\bar{\phi}$,$\bar{\psi}$ to behave
like $\bar{\phi}(\zeta)\approx\kappa_1$ and 
$\bar{\psi}(\zeta)/\zeta\approx\kappa_2$ for large $\zeta$.    
In view of the scaling (\ref{(4.11)}) this makes them consistent
with $\phi(\xi)$,$\psi(\xi)$ finite.
Inserting $\bar{\phi}(\zeta)=\kappa_1$ and 
$\bar{\psi}(\zeta)=\kappa_2\zeta$ into (\ref{(4.13)}) and 
(\ref{(4.15)}), one confirms this ansatz to solve the equations,   
and finds $\kappa_2=-2$ and $\kappa_1=1/6$.
So, again considering (\ref{(4.11)}), the lowest order terms of $\phi$
and $\psi$ as $(\xi-\xi_{tip})$ tends to zero are $\phi=1/6$ and
$\psi=\xi_{tip}/2-2(\xi-\xi_{tip})$. In other words, at $\xi_{tip}$ 
we have the boundary conditions 

\begin{eqnarray}\label{(4.16)}
&& \phi(\xi_{tip})=1/6 \quad, \\
\nonumber
&& \psi(\xi_{tip})=\xi_{tip}/2 \quad. \\
\nonumber
\end{eqnarray}
This boundary condition implies that the asymptotic shape of $h$ 
is a step function of height $|t'|/6$ at the point $z_{tip}(t)$.

It is important to notice that this result is independent of the
particular form of regularization (\ref{(4.4)}),(\ref{(4.5)}) we
have been using. For example any other term involving $h$ leads to
a term $\ell^2\bar{\phi}$ and drops out as $\ell\rightarrow 0$.
Another contribution $\partial_{z'}h$ gives $\bar{\phi}'$ and also does not
contribute as we finally set $\bar{\phi}=\kappa_1$.

It only remains to formulate boundary conditions for $\xi\rightarrow\infty$.
At large distances from the singular point $z_0$ both the interface and 
the velocity field should look the same as before breakup. 
This is the same reasoning that made us construct solutions which far 
away from $z_0$ are static on the time scale of the singularity.
As the width of the singular region shrinks to zero like $|t'|^{1/2}$,
the large body of fluid outside is not able to follow. 
Here it provides us with the mechanism for unique continuation:
For the two solutions to coincide we must require that 

\begin{eqnarray}\label{(4.17)}
&& \lim_{\xi\rightarrow\pm\infty} \phi(\xi)/\xi^2=a_0^{\pm} \quad, \\
\nonumber
&& \lim_{\xi\rightarrow\pm\infty} \psi(\xi)\xi=b_0^{\pm} \quad. \\
\nonumber
\end{eqnarray}

We will see that (\ref{(4.16)}),(\ref{(4.17)}) are all the boundary 
conditions needed to uniquely solve (\ref{(2.24)}),(\ref{(2.25)}) 
after breakup. Since the constants $a_0$ and $b_0$ are different 
for the left and right hand side of the problem, the solutions will
also differ. In particular, the value of $\xi_{tip}$ consistent 
with (\ref{(4.17)}) depends on $a_0$ and $b_0$. The requirement
(\ref{(4.17)}) thus represents the way the properties of the solution 
before breakup are communicated to the solution after breakup. 
Inserting the ansatz 

\begin{eqnarray}\label{(4.18)}
&& \phi=1/6+\phi_1(\xi-\xi_{tip})^{\alpha}+\dots \quad, \\
\nonumber
&& \psi=\xi_{tip}/2-2(\xi-\xi_{tip})+e_0(\xi-\xi_{tip})^{\beta}+\dots \\
\nonumber
\end{eqnarray}
into the similarity equations and balancing leading powers we
find $\alpha=2/5$ and $\beta=7/5$. We therefore try the general 
expansion 

\begin{equation}\label{(4.19)}
\psi=\xi_{tip}/2+(\xi-\xi_{tip})\left[-2+\sum_{i=0}^{\infty}
e_i(\xi-\xi_{tip})^{(2+i)/5}\right] \quad.
\end{equation}

Again, by (\ref{(3.2)}) it is sufficient to consider the expansion of 
$\psi$. The first few coefficients are

\begin{eqnarray}\label{(4.20)}
&& e_0=\frac{60}{7}\phi_1 \quad,\\
\nonumber
&& e_1=0 \quad,\\
\nonumber
&& e_2=-\frac{120}{7}\phi_1^2 \quad,\\
\nonumber
&& e_3=-\frac{5}{72}\xi_{tip} \quad.\\
\nonumber
\end{eqnarray}

We confirmed, by deriving recursion relations for the $e_i$ to
arbitrarily high order, that all coefficients are determined 
by the two free parameters $\xi_{tip}$ and $\phi_1$. Since the 
power series (\ref{(4.19)}) has again a finite radius of 
convergence, all solutions starting from $\xi_{tip}$ are classified 
by just two parameters. But for $t>t_0$ the behavior for 
$|\xi|\rightarrow\infty$ is very different from the situation 
before breakup. We now have $s=-1$, and according to (\ref{(3.9)})
the asymptotic behavior $\phi\sim\xi^2$, $\psi\sim\xi^{-1}$
is {\it stable}. So integrating the similarity equations to
infinity, for every value of $\xi_{tip}$ and $\phi_1$ we will find
a unique value of $\lim_{\xi\rightarrow\infty}\phi(\xi)/\xi^2$
and $\lim_{\xi\rightarrow\infty}\psi(\xi)\xi$. Hence the 
boundary conditions (\ref{(4.17)}) are precisely what is needed
to uniquely fix $\xi_{tip}$ and $\phi_1$, and thereby uniquely 
determining the similarity solution $\phi^-$ and $\psi^-$ after breakup.

Obviously, this has to be done for the left and right hand sides
separately. The left hand side corresponds to a receding neck,
the other is the main drop. We denote the values of $\xi_{tip}$
and $\phi_1$ by $\xi_{neck}$ and $\phi_{neck}$ for the left hand
side, and $\xi_{drop}$ and $\phi_{drop}$ for the right hand side. 
The result
of a numerical calculation of $\phi^-$ and $\psi^-$ can be found
in Figure 4, some of the characteristics of the solution are
listed in Table 2. Specifically, the neck recedes with the 
velocity 

\begin{equation}\label{(4.21)}
v_{neck}=\frac{\xi_{neck}}{2} |t'|^{-1/2} \quad,
\end{equation}
where $\xi_{neck}/2\approx 8.7$. Unfortunately, on the scale of
Figure 4 it is hard to see any deviations from a flat interface 
for the drop. Figure 5 below will give a better idea of how the
drop is left distorted after breakup. 

It should be appreciated that the unique continuation does not 
follow from the asymptotic equations (\ref{(2.24)}),(\ref{(2.25)})
alone. Rather, we needed to invoke regularity for $|t'|\neq 0$ to
derive the boundary condition $\phi^-(\xi_{tip})=1/6$. Indeed,
(\ref{(2.24)}) and (\ref{(2.25)}) with $s=-1$ would allow for 
an infinity of solutions, one for each value of $\phi(\xi_{tip})$.

\section{Discussion}
We have shown that the motion of a Navier-Stokes fluid
close to the time of breakup is described by self-similar 
solutions. The corresponding scaling functions, before and 
after the breakup, are solutions to a set of ordinary differential
equations. For the solutions to be consistent, both away from
the singular point and at the receding tip after breakup, 
boundary conditions have to be imposed. They lead to unique 
solutions of the similarity equations. This means solutions 
to the Navier-Stokes equation close to the singularity are 
predicted without adjustable parameters, and independent of 
boundary or initial conditions. It is quite
instructive to plot the predicted interface of a real fluid 
at constant time intervals before and after the singularity.
Since $\ell_{\nu}$ and $t_{\nu}$ are almost on molecular 
scales for water \cite{/ED/}, we take a mixture of
glycerol and ethanol as a reference fluid, for which 
$\ell_{\nu}=72\mu m$ and $t_{\nu}=114\mu s$. This is large
enough for experiments by optical means to be feasible.
Measurements of the velocity field are also possible \cite{/M/}.
Figure 5 shows three profiles, each $46\mu s$ apart, before the 
singularity (a), and after the singularity (b). This corresponds
to $|t'|=1$, $0.55$, and $0.1$. In particular, there is no freedom
in the spatial scale of this Figure. The same graph should 
apply regardless of boundary conditions.

Before breakup, one can clearly distinguish a very slender neck,
and the steep front of the adjoining drop. As the neck becomes
thinner, the minimum moves towards the drop, making the interface even 
steeper. The greatest relative changes in the diameter occur 
near the minimum, far away the interface is practically static.
As one comes closer to the singularity, the size of the ``active''
region, which is still changing, becomes smaller and smaller.

After breakup, the neck snaps back very rapidly, forming a 
sharp front at the end. For $|t'|=1$, higher order corrections
in $|t'|$ will probably be already important, and the end will
look more rounded. As seen in Figure 4, there is also some 
fluid accumulating at the end in the asymptotic solutions, 
but this cannot be seen on the scale of Figure 5. The small
protrusion on the drop, left by the breakup, quickly relaxes
to an almost flat interface.

The asymmetry of the breakup was already noticed in experiments
\cite{/GY/,/PSS/}. However, one must be careful not to apply 
our results to those experiments directly, since they are on 
length and time scales $z'\gg 1$,$|t'|\gg 1$, far away from the 
asymptotic behavior. Still our similarity solutions could play 
a crucial role for the shape selection even in this ``inviscid''
regime, since all solution must ultimately match onto the 
asymptotic behavior. Clearly, an extension of our theory 
to the almost inviscid regime seems highly desirable. 

Recently, an experimental study of drop formation in a highly 
viscous fluid has been reported \cite{/SBN/}. Qualitatively,
the shape of the interface adjoining the primary drop agrees 
well with Figure 5. Also, the length and time scales of the 
similarity solution, as given by the present theory, 
have been used to analyze the data and
are found to be consistent with experiment. Unfortunately,
at the times shown in Figure 2A and Figure 2B of \cite{/SBN/}, the
straining due to the falling drop is still appreciable compared 
with the scales of the similarity solution. Also, there is no
independent measurement of $t'$ available, which makes a meaningful
comparison with theory difficult at present. We will discuss the 
process of repeated necking, reported in \cite{/SBN/}, below.

Extensive experiments with high-speed jets, where gravity is 
irrelevant, are also in progress \cite{/K/}. The stroboscopical 
method employed for example in \cite{/BHK/} allows to determine
$t'$ independently, so comparison with theory can be made 
without adjustable parameters. Preliminary results show nice
agreement with theory before breakup. After breakup a quantitative
comparison with theory is not yet possible, due to air drag on
the rapidly receding neck, whose effect is not yet included 
in the equations.

Therefore, we will use numerical simulations for a detailed 
comparison with theory.
In particular, we would like to verify the prediction of the
theory that the same similarity solutions are always approached,
independent of boundary or initial conditions.
Indeed, some simulations have already 
been performed on the breakup of a Navier-Stokes fluid \cite{/SE/},
but they are not sufficiently close enough to the singularity to 
allow for a meaningful comparison. This is because in the asymptotic
region Navier-Stokes computations become prohibitively expensive.
To make simulations feasible, one has to resort to approximations.    

As model equations, we take the generalized form of the asymptotic
equations (\ref{(4.1)}) and (\ref{(4.2)}). Extensive simulations
of this system before breakup were already reported in \cite{/ED/},
\cite{/E/}, and \cite{/SBN/}. The equations read 

\begin{equation}\label{(5.1)}
\partial_t v_0 + v_0 \partial_z v_0 = -\frac{\gamma}{\rho} \partial_z p
+ 3\nu \frac{\partial_z\left[(\partial_z v_0) H^2\right]}{H^2} - g \quad,  
\end{equation}

\begin{equation}\label{(5.2)}
\partial_t H + v_0 \partial_z H = -(\partial_z v_0) H/2 \quad, 
\end{equation}
where

\begin{equation}\label{(5.3)}
p = \frac{1}{H(1 + (\partial_z H)^2)^{1/2}} - 
\frac{\partial_z^2 H}{(1 + (\partial_z H)^2)^{3/2}} \quad.
\end{equation}
So apart from the asymptotic terms already contained in
(\ref{(2.18)}) and (\ref{(2.19)}), (\ref{(5.3)}) contains the
exact expression for the mean curvature of a body of revolution.
The system (\ref{(5.1)})-(\ref{(5.3)}) was supplemented with 
two types of boundary conditions \cite{/ED/}:

In the ``jet geometry'' we fix the values of $H$ and $v_0$ at
two fixed points $z_+$ and $z_-$ : 

\begin{equation}\label{(5.4)}
H(z_{\pm},t) = H_{\pm}(t) \quad,
\end{equation}

\begin{equation}\label{(5.5)}
v_0(z_{\pm},t) = v_{\pm}(t) \quad.
\end{equation}
Hence here we envision a jet of length $z_+-z_-$ with nozzle radius 
$H_+=H_-\equiv r_0$ and speed $v_+=v_-\equiv V$. At some point in time
a small perturbation is applied to the speed $v_-$ at the nozzle and
the jet breaks up according to the Rayleigh instability. The jet speed is so 
high that gravitational effects can be neglected, and thus $g = 0$.

In the ``drop geometry'' fluid is released slowly from a tap. 
Thus at the opening of the tap,$z_-$ say, boundary conditions 
(\ref{(5.4)}) and (\ref{(5.5)}) hold, while at the lower end of 
the drop the boundary moves with the fluid. This means we have

\begin{equation}\label{(5.6)}
H(z_+(t),t) = 0
\end{equation}
and

\begin{equation}\label{(5.7)}
v_0(z_+(t),t) = \partial_t z_+(t) \quad.
\end{equation}
In this experimental situation gravity is of course important,
as initially gravitational and surface tension forces are
balanced, and the drop assumes an equilibrium shape \cite{/MW/}. These 
shapes are reproduced exactly by the stationary solutions of
(\ref{(5.1)})-(\ref{(5.3)}). Eventually, gravity overcomes 
surfaces tension and the drop falls and subsequently pinches off.

The implementation of boundary conditions as well as the 
numerical procedure is explained in detail in \cite{/ED/}.
In \cite{/ED/} and \cite{/SBN/} simulations of (\ref{(5.1)})-(\ref{(5.3)})
have been used to reproduce experimental interface shapes 
both for high and low viscosity fluids in different 
geometries. In particular in the case of a slowly dripping 
tap \cite{/PSS/}, both boundary and initial conditions are known 
and comparison with experiment can be done without adjustable
parameters. Thus the excellent agreement between simulation 
and the experimental shape of a falling drop at the pinch point
\cite{/ED/} seems highly significant. Therefore we are confident 
that (\ref{(5.1)})-(\ref{(5.3)}) represents a good approximation to 
the Navier-Stokes equation not only close to the pinch point,
but also for earlier times and including the crossover to the boundary.

We have performed systematic tests of the predictions
of the present theory, in particular investigating the independence
of the singular behavior near break- off from boundary conditions.
For all runs, both in the jet and the drop geometry, and independent 
of the nozzle or tap diameter and of the viscosity, we always found 
the flow to converge onto the similarity solution predicted by
the present theory. 

Figure 6 shows this convergence for a typical run in the jet 
geometry. The nozzle diameter is 100 in units of $\ell_{\nu}$.
The solution near the singularity has been converted to similarity 
variables, thus giving $\phi(\xi)$ and $\psi(\xi)$ using the 
transformation (\ref{(2.23)}). Shown is the predicted similarity 
solution as a solid line, and the computed solution at times 
$|t'| = 0.39$, $0.13$, $0.043$, and $0.014$, represented by 
dashed, chain-dashed, dot-dashed, and dotted lines. It can clearly 
be seen that the range of validity of the similarity solution 
{\it expands} like $|t'|^{-1/2}$ in the similarity variable $\xi$.
This means there is a {\it fixed} region in $z'$ where the 
similarity theory applies, in agreement with the statements of
Section 3. At the boundary of this region, 
the slope $\partial_{z'} h$ becomes of order unity, and the 
expansion in orders of $\epsilon$ breaks down.

The motion shown in Figure 6 occurs on scales widely separated 
 from those imposed by the boundary conditions. The time distance 
from the onset of the linear instability to the singularity is
$t_0/t_{\nu} = 32284$, much larger than the relevant $t'$. 
Similarly, the nozzle diameter, converted to the similarity 
variable $\xi$, is $\xi = 157$, $272$, $473$, and $829$, for the
times shown. Clearly the motion near the singularity has 
become independent of these imposed length and time scales.
The same will happen for {\it any} boundary condition, as 
both the typical time and length scale shrinks to zero near 
the singularity.

Next we test the convergence onto the similarity solution 
after breakup. Since there is a moving tip, we modify
the viscous term in (\ref{(5.1)}) to regularize the tip. 
The force balance now reads 

\begin{equation}\label{(5.8)}
\partial_t v_0 + v_0 \partial_z v_0 = -\frac{\gamma}{\rho} \partial_z p
+ 3\nu \frac{\partial_z\left[(\partial_z v_0) 
H^2/(1 + a(\partial_z H)^2)\right]}{H^2} - g \quad,  
\end{equation}
where $a$ is a free constant. By varying $a$ we can test our
prediction that the shape of the interface after breakup does not
depend on the regularization employed.

To produce an initial condition after breakup, we take a simulation
before breakup, which has progressed to a time distance of $|t'| = 10^{-4}$
from the singularity. Then we cut the solution at the minimum 
and interpolate $H$ to zero with a polynomial, so as to keep the
highest derivatives smooth. This we take as the new initial condition 
after breakup and let the solution evolve under (\ref{(5.8)}), 
(\ref{(5.2)}), and (\ref{(5.3)}). For a wide range of values of 
$a$ in (\ref{(5.8)}), we always find the solution to converge 
onto the similarity form found in Section 4. Figure 7 illustrates
this convergence for a run which has the same boundary conditions
and material parameters as the one shown in Figure 6 before breakup. 
The constant $a$ was chosen to be $1$. Again, solutions were 
converted to similarity variables. The full line represents
the predicted similarity solution, the dot-dashed and the dashed
lines show the numerical simulations for $|t'| = 0.006$ and 
$|t'| = 0.06$ after the singularity. The dotted line is the 
similarity solution before breakup, shown for comparison. It 
can clearly be seen that after the solution has been cut in two halves
it rapidly converges onto the predicted similarity form. 
This is independent of both the regularizing term in (\ref{(5.8)}) 
and the procedure by which the solution is cut.

Hence both before and after the singularity, we have always observed 
convergence onto the similarity solutions if $|t'|$ is small. 
Still it would be very useful to have a better mathematical 
understanding of the approach of the similarity solution for
the full Navier-Stokes dynamics. Even for the simplified model
equations (\ref{(5.1)})-(\ref{(5.3)}) the convergence we found 
numerically is far from being a trivial result, as there are
higher order derivative terms like $\partial_z^3 H$ coming from the 
pressure. In principle, although these terms are multiplied
by a small number $t'$ close to the singularity, they could 
make a singular perturbation, which changes the asymptotics. 
However, it is well beyond the scope of this paper to explore 
these questions in detail, so at present we have to rely on 
the ample numerical evidence.

Another important question is the stability of the similarity
solution to small perturbations. This has been studied in the 
framework of the asymptotic equations (\ref{(2.18)}),(\ref{(2.19)})
in \cite{/BSN/}, both numerically and analytically. The result 
is that the similarity solutions are linearly stable as expected,
since there are observed numerically. On the other hand, 
they are unstable to {\it finite amplitude perturbations} of 
wavelength comparable to the minimum radius of the fluid neck.

As soon as a perturbation is large enough, it will start to 
grow and eventually forms a new similarity solution with its 
own $z_0$ and $t_0$. For any finite number of such perturbations,
the singularities are separated in space and the present theory
strictly applies. However, if one explicitly adds an external
white noise source to the Navier-Stokes equation, perturbations 
are introduced on all time scales arbitrarily close to the 
singularity. This allows for the appearance of a ``rough'' 
interface as described in \cite{/SBN/} and \cite{/BSN/}, 
consisting of an infinity of interacting similarity solutions.
Locally, the form of each of those solutions, seen as ``necks''
in experiment, is consistent with the present theory. 

In \cite{/BSN/}, a threshold length scale 

\begin{equation}\label{(5.9)}
\ell_{thres} \sim \ell_{\nu}\left(\frac{\ell_T}{\ell_{\nu}}\right)^{2/5}
\end{equation}
was identified, below which thermal fluctuations become important.
The relevant thermal length scale for surface perturbations is
$\ell_T = (k_BT/\gamma)^{1/2}$. For a mixture of 85\% glycerol and 
15\% water $\ell_{thres}$ is  
$1 \mu m$. Thus, in the presence of thermal fluctuations,
the microscopic length scale $\ell_{micro}$ introduced in 
Section 3 may be replaced by $\ell_{thres}$: for $H_{min}$ larger 
than $\ell_{thres}$ the Navier-Stokes equation is applicable,
on smaller scales the equations are inherently stochastic.

An obvious benefit one expects from the universality found in the 
present paper is the 
unique continuation of Navier-Stokes {\it simulations} through the
singularity. One slight problem lies in the nonanalyticity 
of $\phi^-$ and $\psi^-$ if one wants to use similarity solutions
as new initial conditions after breakup. Although the pressure
itself would be finite, pressure gradients and the viscous term
would be infinite at the tip. Thus it is better to use regularized
similarity functions, where a small but finite cut-off parameter 
$\ell$ has been introduced. The resulting initial conditions
for the new Navier-Stokes problem after the singularity would 
be arbitrarily close to the similarity form, but still finite
at the tip.

In conclusion, we have shown that the Navier-Stokes equation 
carries us through the bifurcation point where at first it seems 
meaningless. As usual, classical hydrodynamic theory has a much
wider range of applicability than purely microscopic considerations 
would tell us. 

\acknowledgements
I am grateful to a great number of people, for discussions 
and all sorts of help and encouragement. In particular
Hartwig Brand, Michael Brenner, Todd Dupont, Greg Forest,
Siegfried Grossmann, Leo Kadanoff, Michael Tabor, and
Stephane Zaleski.

\begin{table}
\begin{tabular}{ccccccccc}
$\xi_0$ & $\phi_0$ & $\xi_{min}$ & $\phi_{min}$ & $\psi_{max}$ & $a_0^+$
& $b_0^+$ & $a_0^-$ & $b_0^-$ \\ \hline
$-1.5699$ & $0.030432$ & $-1.6024$ & $0.030426$ & $-3.066$ & $4.635$ & 
$0.0723$ & $6.047\times 10^{-4}$ & $57.043$ \\
\end{tabular} 
\caption{
Some characteristics of the similarity functions $\phi^+$,$\psi^+$
before breakup. The symbols $\xi_0$ and $\phi_0$ stand for the 
position of the stagnation point, where the fluid is at rest in 
the frame of reference of the interface, and the radius of the 
interface at that point. The minimum value of $\phi^+$ is $\phi_{min}$,
and $\xi_{min}$ is its position. The function $\psi^+$ reaches
a maximum value of $\psi_{max}$. The numbers $a_0^{\pm}$ and
$b_0^{\pm}$ stand for the limits 
$\lim_{\xi\rightarrow\pm\infty} \phi^+(\xi)/\xi^2$ and
$\lim_{\xi\rightarrow\pm\infty} \psi^+(\xi)\xi$, respectively.
All numbers are accurate to the decimal places shown.
} 
\end{table}

\begin{table}
\begin{tabular}{cccc}
$\xi_{neck}$ & $\phi_{neck}$ & $\xi_{drop}$ & $\phi_{drop}$ \\ \hline
$17.452$ & $0.06183$ & $0.4476$ & $0.6180$ \\
\end{tabular} 
\caption{
Characteristics of the similarity functions $\phi^-$,$\psi^-$
after breakup. The tip position of the left hand, or neck side 
is $\xi_{neck}$, and the expansion coefficient $\phi_1$, cf. 
(\protect\ref{(4.18)}), is $\phi_{neck}$. Correspondingly,
$\xi_{drop}$ and $\phi_{drop}$ uniquely determine the ``drop''
side of $\phi^-$ and $\psi^-$. The values of $a_0^{\pm}$ and
$b_0^{\pm}$ are the same as before breakup, cf. Table 1.
} 
\end{table}

\begin{figure}
\caption{
A sketch of the flow geometry investigated in the present paper.
The radius or ``height'' of the free surface at a point $z$ on
the axis of symmetry is $H(z)$. The velocity field inside the fluid
is ${\bf v}(z,r) = v_z(z,r) {\bf e}_z + v_r(z,r) {\bf e}_r$.
} 
\label{fig1}
\end{figure}

\begin{figure}
\caption{
A plot of the similarity functions $\phi^+$, (a), and $\psi^+$, (b),
before breakup. 
Note the strong asymmetry. 
} 
\label{fig2}
\end{figure}

\begin{figure}
\caption{
A cartoon of a receding tip after breakup. The position of the
tip is $z'_{tip}(t)$.
}
\label{fig3}
\end{figure}

\begin{figure}
\caption{
The similarity functions $\phi^-$, (a), and $\psi^-$, (b),
which are unique continuations of $\phi^+$ and $\psi^+$ to
times greater than $t_0$. The asymptotic behavior for 
$\xi\rightarrow\pm\infty$ is by definition the same as before 
breakup. On the left is the rapidly receding ``neck'' part
of the solution, on the other side is the drop. The points at
$\xi_{neck}$ and $\xi_{drop}$, from where the interface is 
plane, are marked by diamonds.
}
\label{fig4}
\end{figure}

\begin{figure}
\caption{
The breakup of a mixture of 5 parts of glycerol in 
4 parts of ethanol, as calculated
from the similarity solutions. Part (a)
shows three profiles before breakup, in time distances of $46 \mu s$,
corresponding to $|t'|=1$,$0.55$, and $0.1$. In part (b) the 
same is shown after breakup.
}
\label{fig5}
\end{figure}

\begin{figure}
\caption{
Simulation of (\protect\ref{(5.1)})-(\protect\ref{(5.3)}) in the 
jet geometry. The profiles close to pinch-off were converted to 
similarity variables. The full line is the prediction of the 
present theory; the dashed, chain-dashed, dot-dashed, and dotted
lines represent the simulation at $|t'| = 0.39$, $0.13$, $0.043$,
and $0.014$. The inset contains a blowup of the central region
with only the latest time, $|t'| = 0.014$.  
} 
\label{fig6}
\end{figure}

\begin{figure}
\caption{
The approach of the similarity function $\phi^-$ by the solution
of (\protect\ref{(5.8)}),(\protect\ref{(5.2)}) and
(\protect\ref{(5.3)}) in the jet geometry,
transformed to similarity variables.
The fluid neck is severed at $|t'|=10^{-4}$ before breakup. 
The full line is $\phi^-$, the dotted line the solution before breakup.
The dot-dashed
and the dashed lines show the simulation at $|t'|=0.006$ and
$|t'|=0.06$, respectively.
}
\label{fig7}
\end{figure}

\newpage
\begin{figure}
  {\Huge Figure 1:}
  \begin{center}
    \leavevmode
    \epsfsize=0.9 \textwidth
    \epsffile{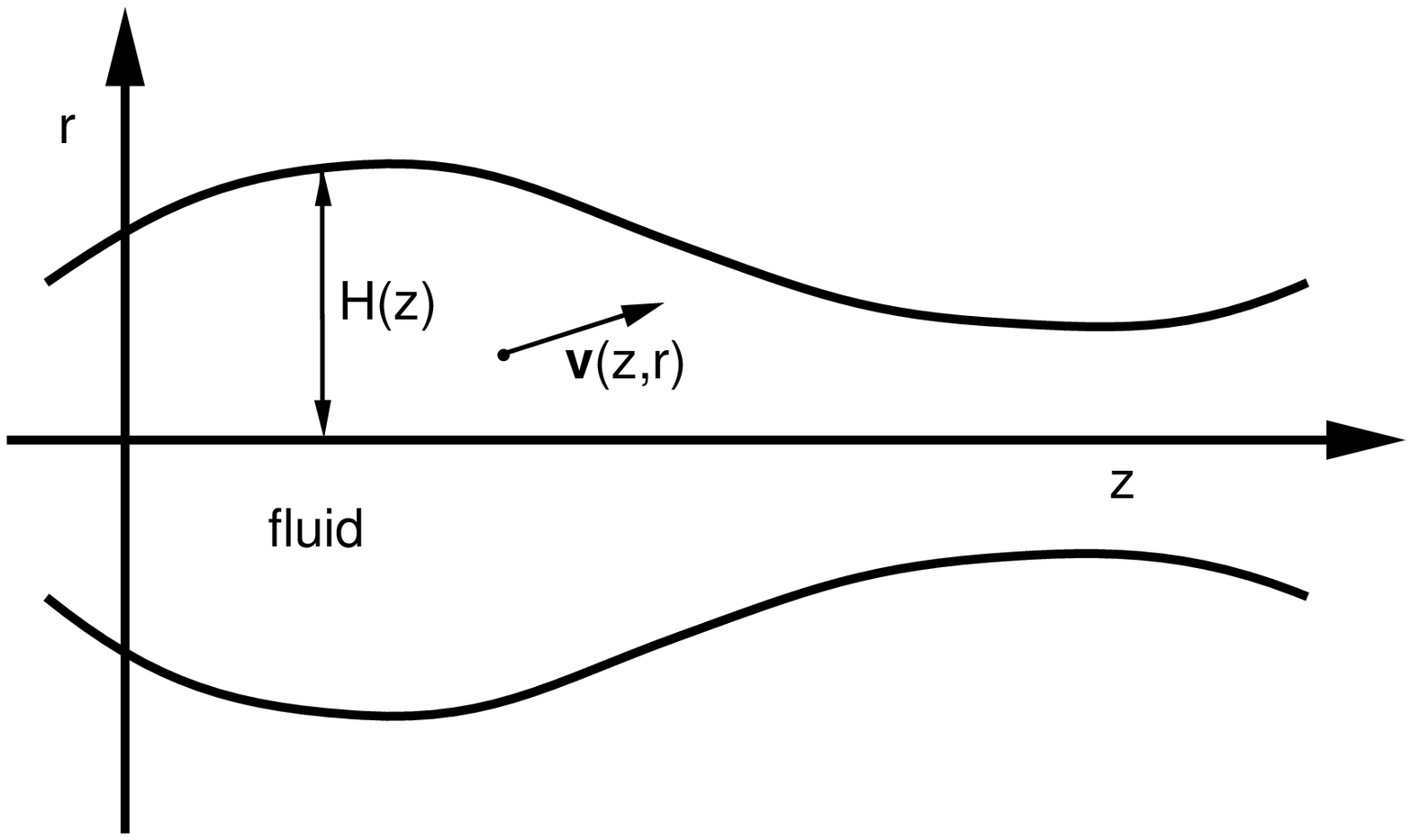}
  \end{center}
\end{figure}

\newpage
\begin{figure}
  {\Huge Figure 2a:}
  \begin{center}
    \leavevmode
    \epsfsize=0.9 \textwidth
    \epsffile{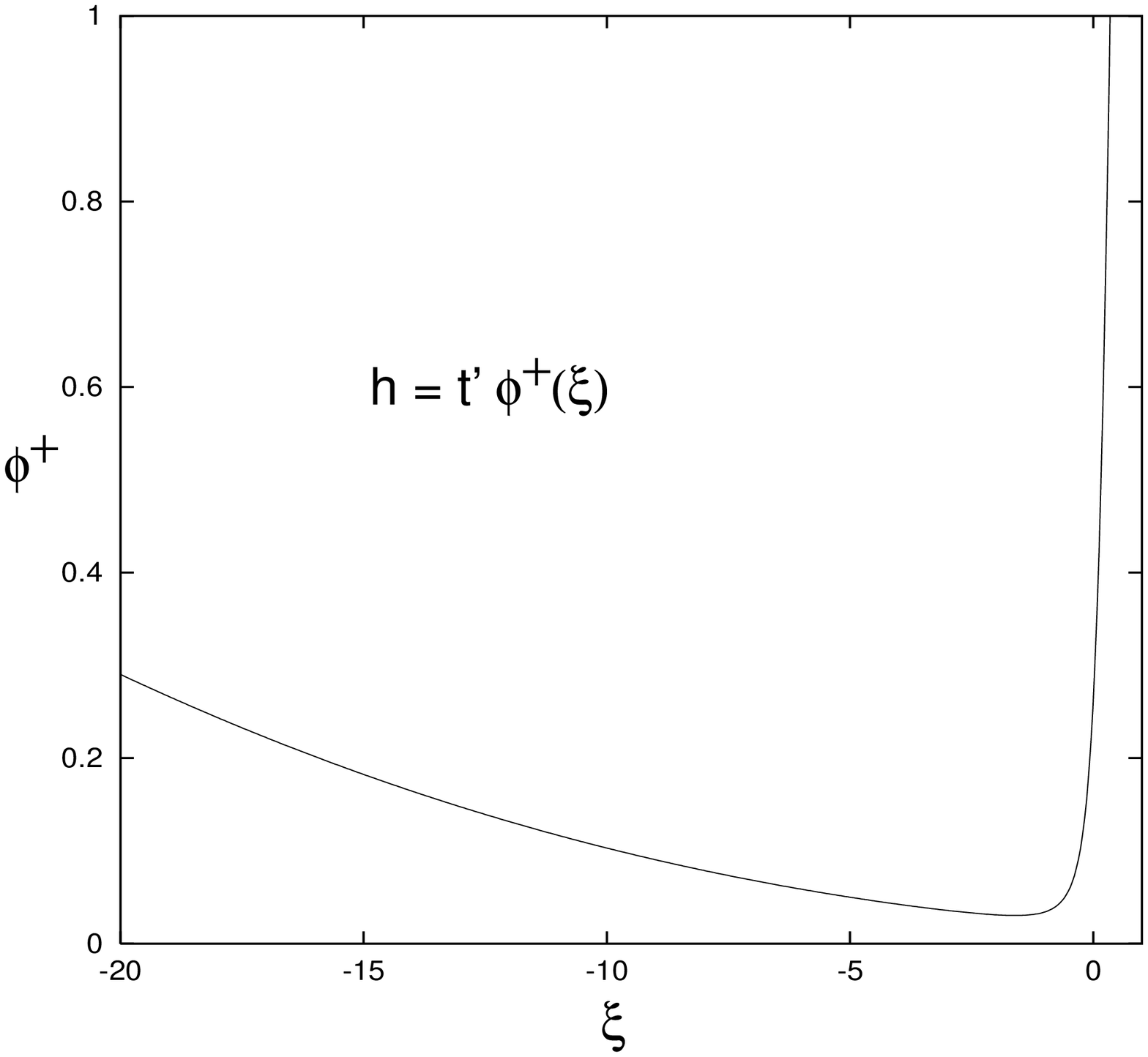}
  \end{center}
\end{figure}

\newpage
\begin{figure}
  {\Huge Figure 2b:}
  \begin{center}
    \leavevmode
    \epsfsize=0.9 \textwidth
    \epsffile{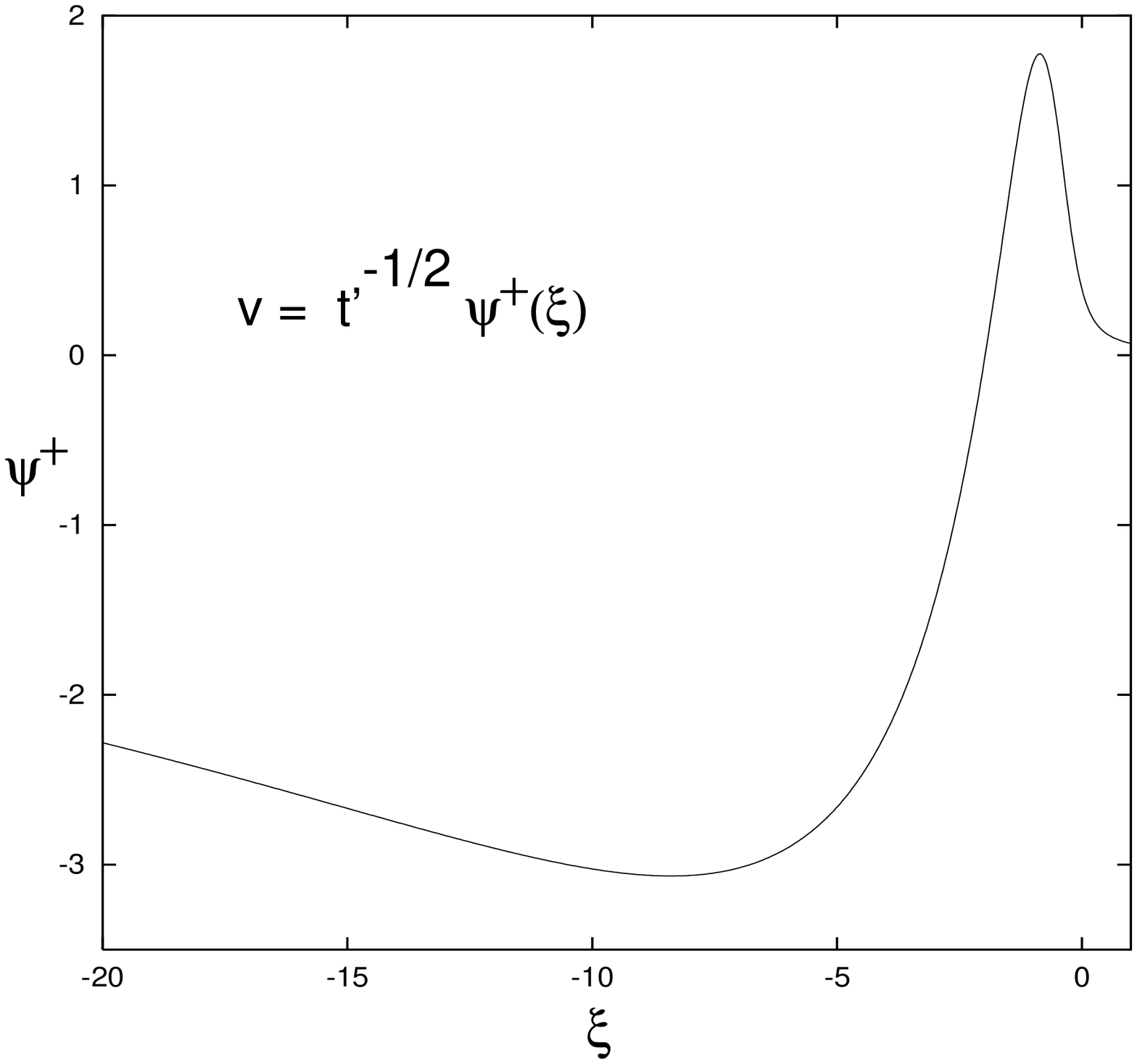}
  \end{center}
\end{figure}

\newpage
\begin{figure}
  {\Huge Figure 3:}
  \begin{center}
    \leavevmode
    \epsfsize=0.9 \textwidth
    \epsffile{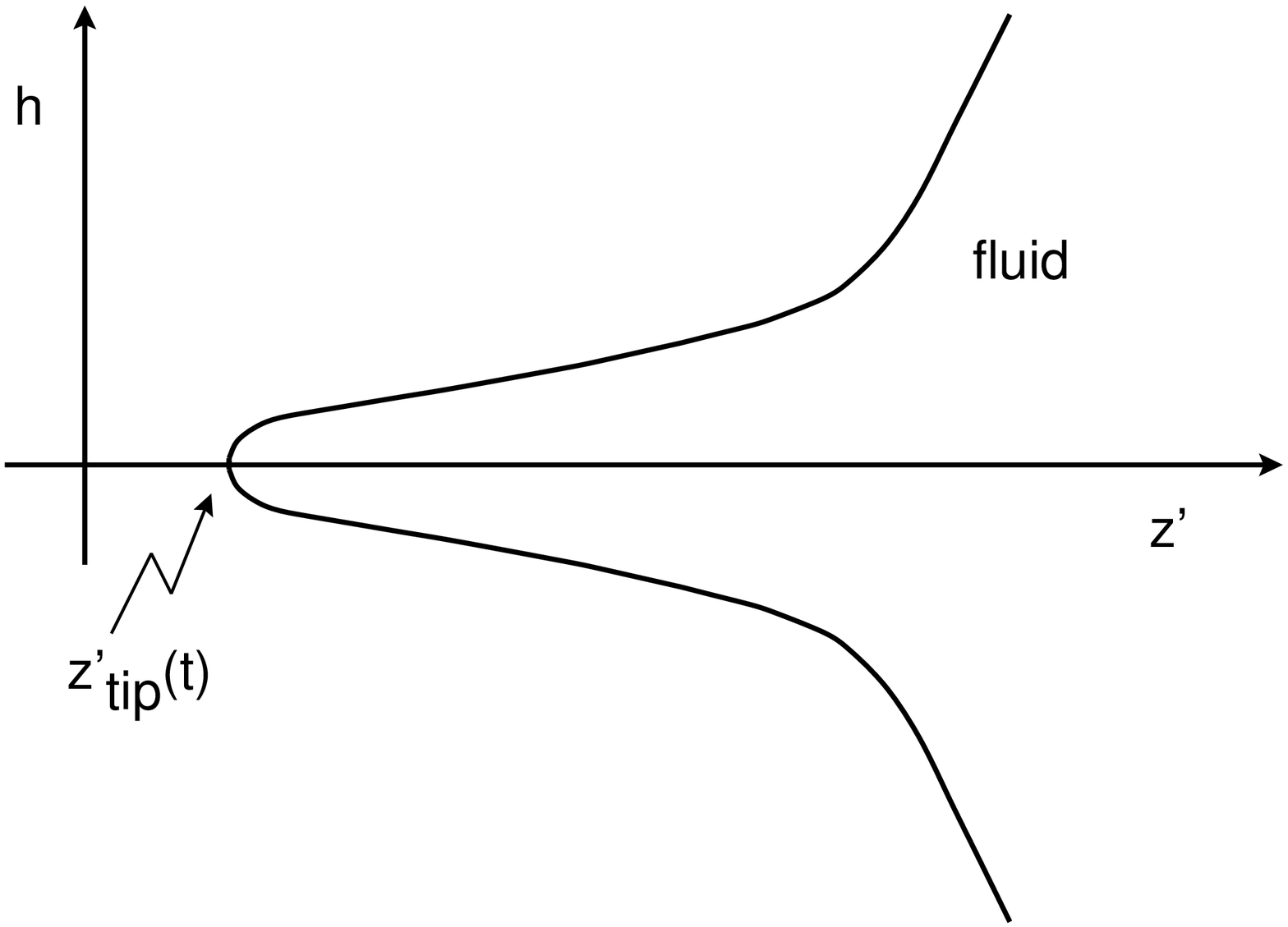}
  \end{center}
\end{figure}

\newpage
\begin{figure}
  {\Huge Figure 4a:}
  \begin{center}
    \leavevmode
    \epsfsize=0.9 \textwidth
    \epsffile{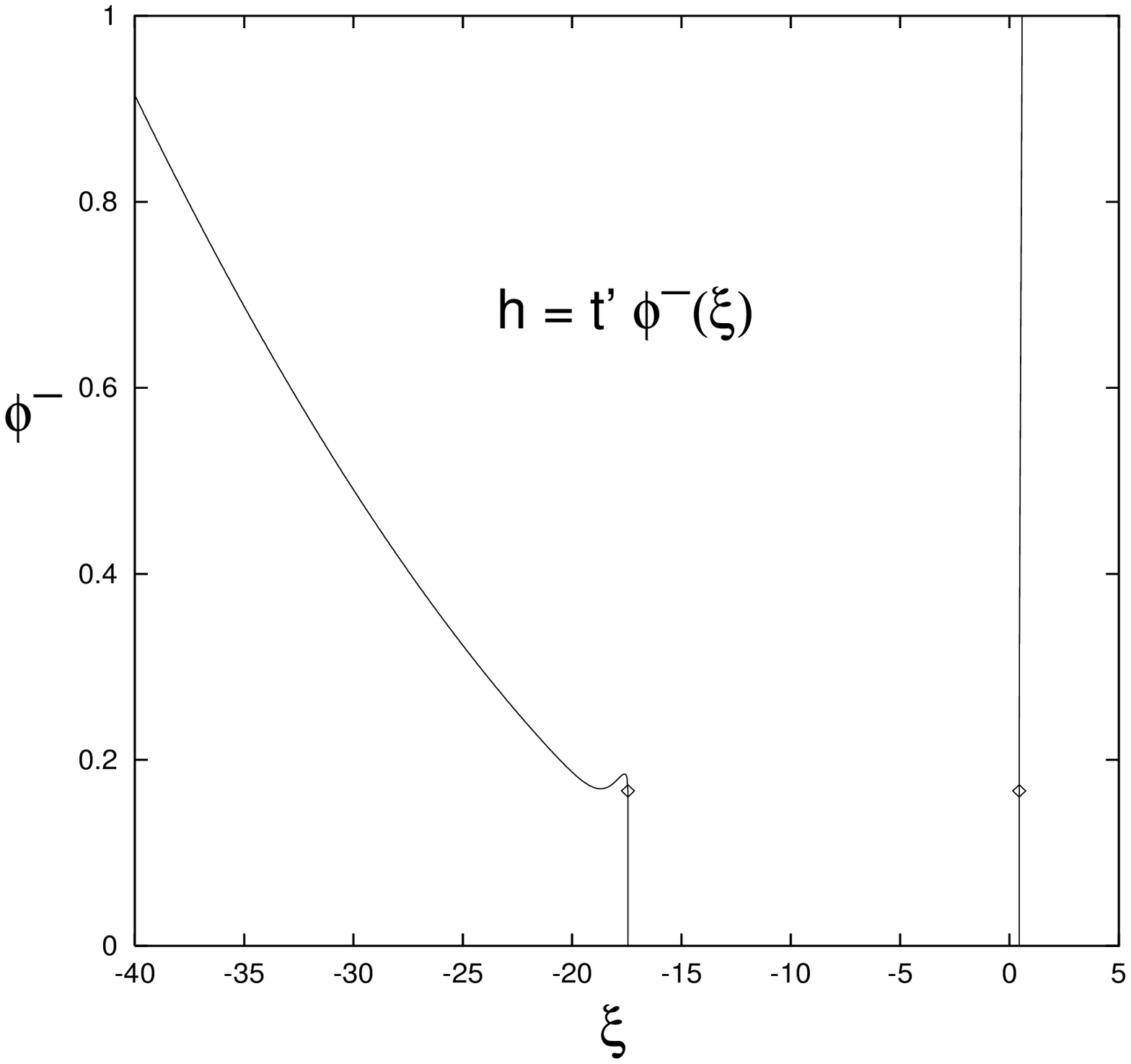}
  \end{center}
\end{figure}

\newpage
\begin{figure}
  {\Huge Figure 4b:}
  \begin{center}
    \leavevmode
    \epsfsize=0.9 \textwidth
    \epsffile{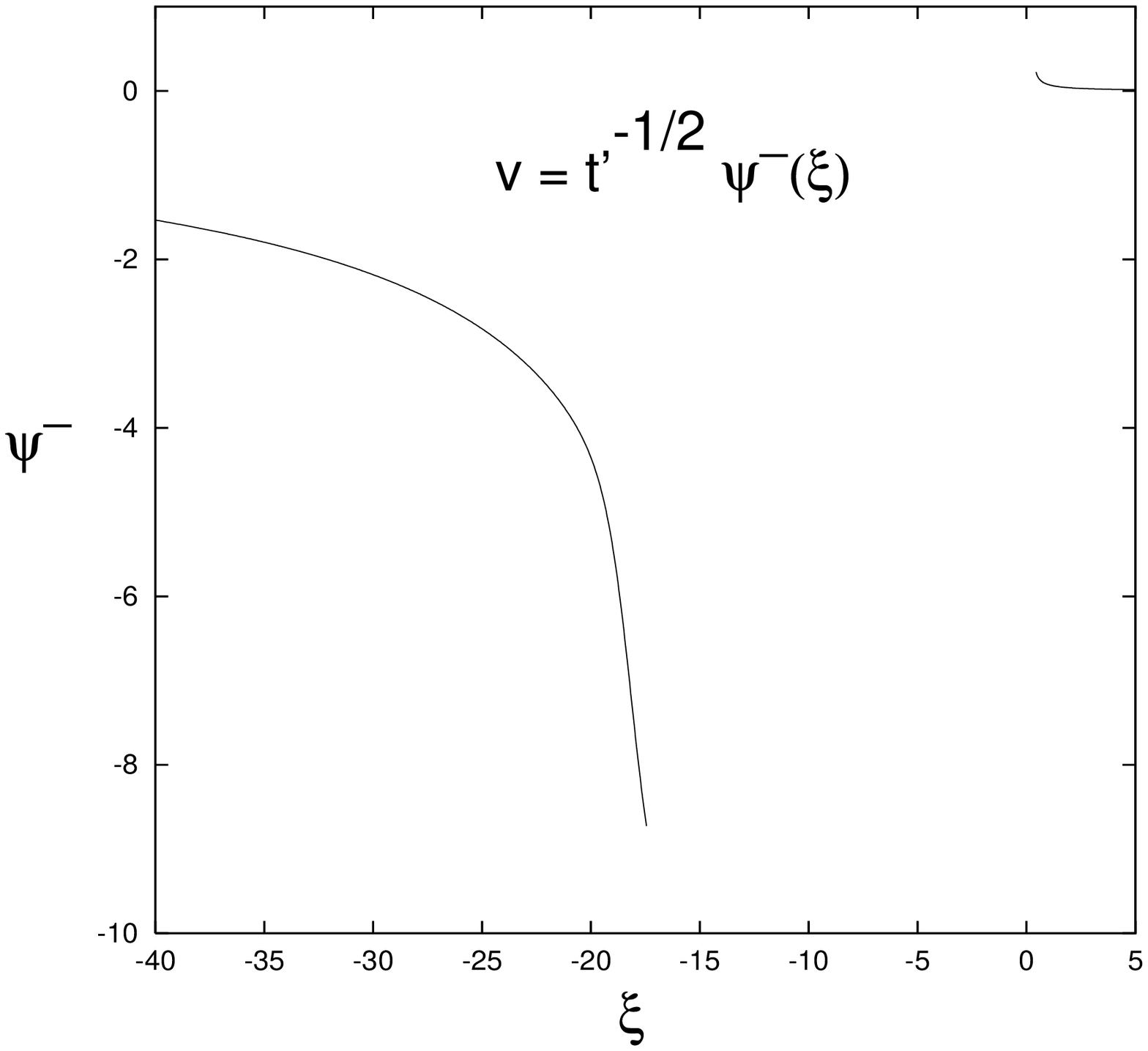}
  \end{center}
\end{figure}

\newpage
\begin{figure}
  {\Huge Figure 5a:}
  \begin{center}
    \leavevmode
    \epsfsize=0.9 \textwidth
    \epsffile{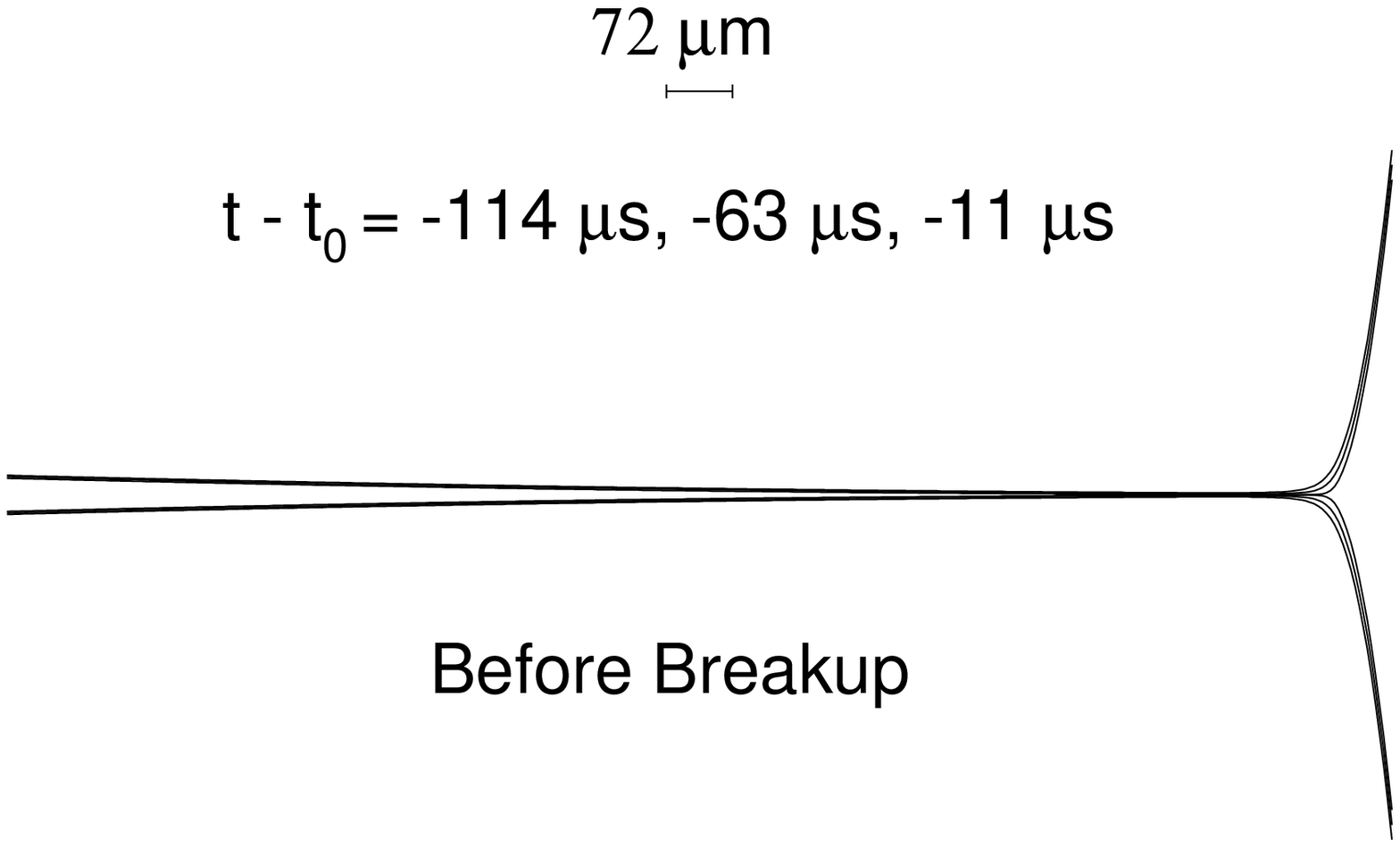}
  \end{center}
\end{figure}

\newpage
\begin{figure}
  {\Huge Figure 5b:}
  \begin{center}
    \leavevmode
    \epsfsize=1\textwidth
    \epsffile{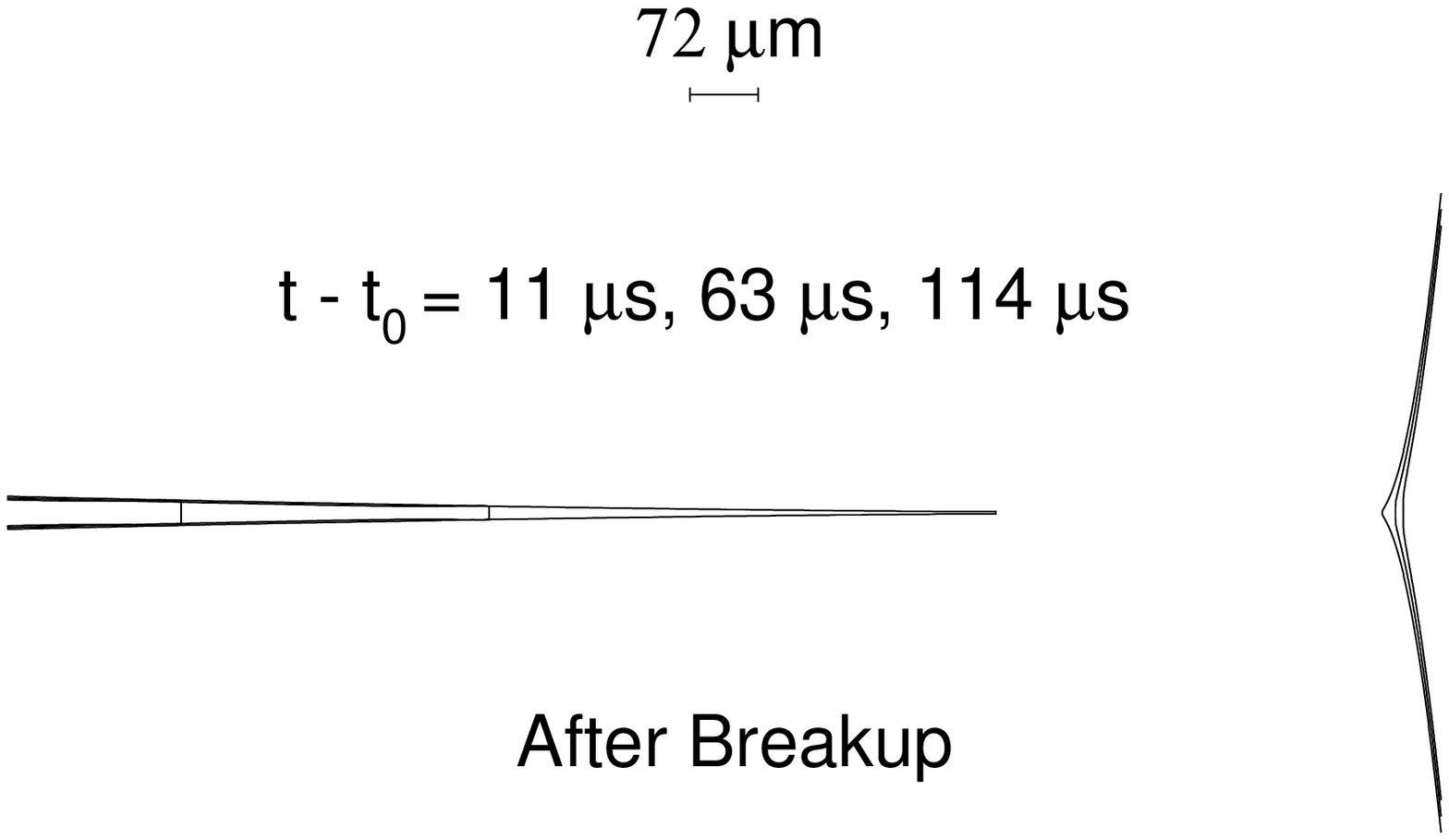}
  \end{center}
\end{figure}

\newpage
\begin{figure}
  {\Huge Figure 6a:}
  \begin{center}
    \leavevmode
    \epsfsize=0.9 \textwidth
    \epsffile{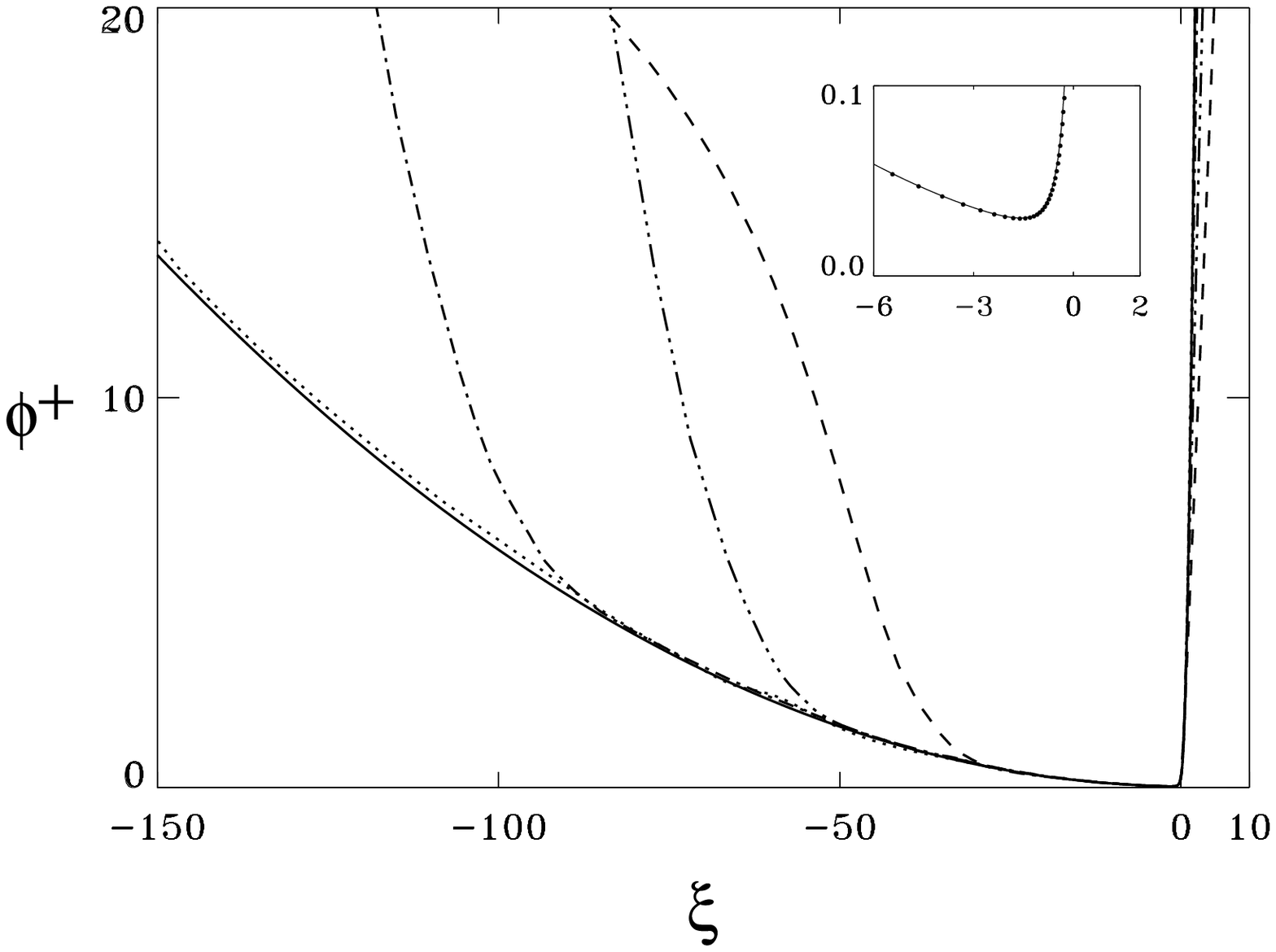}
  \end{center}
\end{figure}

\newpage
\begin{figure}
  {\Huge Figure 6b:}
  \begin{center}
    \leavevmode
    \epsfsize=0.9 \textwidth
    \epsffile{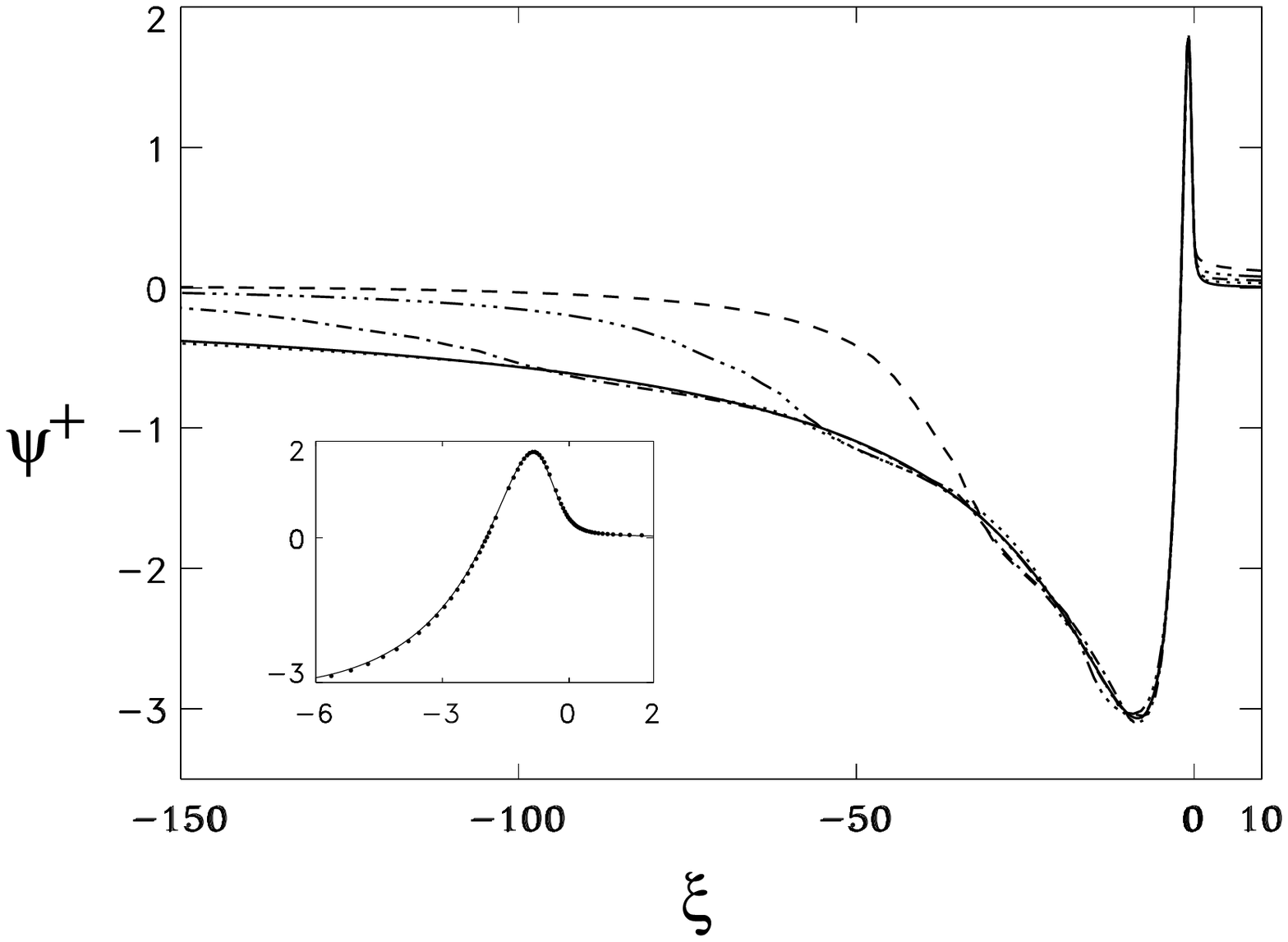}
  \end{center}
\end{figure}

\newpage
\begin{figure}
  {\Huge Figure 7:}
  \begin{center}
    \leavevmode
    \epsfsize=0.9 \textwidth
    \epsffile{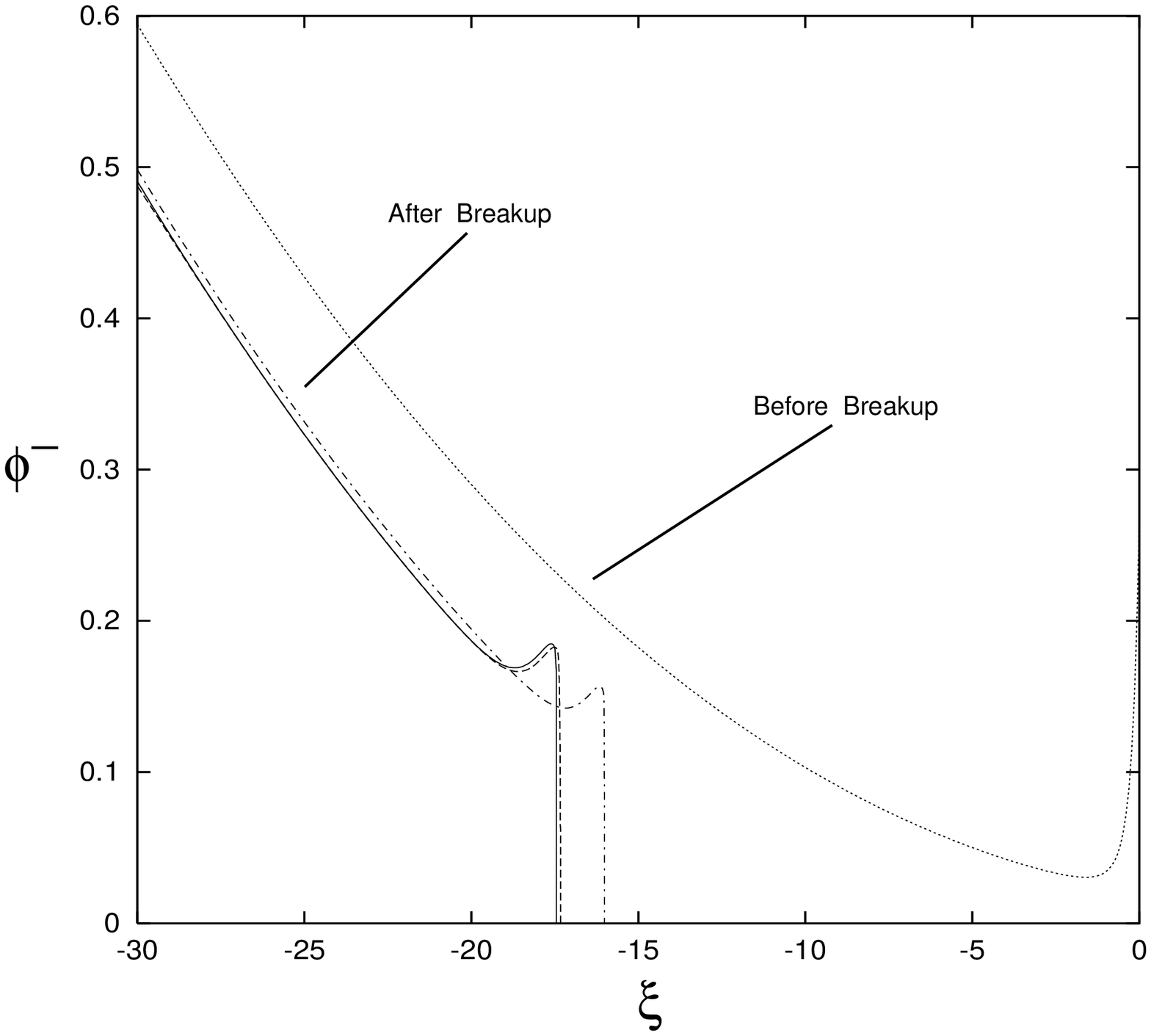}
  \end{center}
\end{figure}

\end{document}